\documentclass[12pt,cls,onecolumn]{IEEEtran}
\usepackage{graphicx,amsmath,amssymb,epsfig, amsfonts, cite, latexsym, cuted, multicol, multirow, subfigure, stfloats, array, tabularx}
\usepackage{graphicx,psfig,amsmath,amssymb,mathrsfs,epsf}
\usepackage{subeqnarray,tabularx}
\usepackage{color}
\usepackage{setspace}
\usepackage{anysize}

\begin{document}

\title{GreenInfra: Capacity of Large-Scale Hybrid Networks With Cost-Effective Infrastructure}
\author{\large Cheol~Jeong,~\IEEEmembership{Member,~IEEE}
        and~Won-Yong~Shin,~\IEEEmembership{Member,~IEEE}
\\
\thanks{This research was supported by the Basic Science Research
Program through the National Research Foundation of Korea (NRF)
funded by the Ministry of Education (2014R1A1A2054577). This paper
was presented in part at the 2013 IEEE International Symposium on
Information Theory, Istanbul, Turkey, July 2013.}
\thanks{C. Jeong is with the DMC R\&D Center, Samsung Electronics, Suwon 443-742, Republic of Korea (E-mail:
cheol.jeong@ieee.org).}
\thanks{W.-Y. Shin (corresponding author) is with the Department of Computer Science and
Engineering, Dankook University, Yongin 448-701, Republic of Korea
(E-mail: wyshin@dankook.ac.kr).}
} \maketitle


\markboth{IEEE Journal on Selected Areas in Communications}%
{Under Review for Possible Publication}


\newtheorem{definition}{Definition}
\newtheorem{theorem}{Theorem}
\newtheorem{lemma}{Lemma}
\newtheorem{example}{Example}
\newtheorem{corollary}{Corollary}
\newtheorem{proposition}{Proposition}
\newtheorem{conjecture}{Conjecture}
\newtheorem{remark}{Remark}

\def \diag{\operatornamewithlimits{diag}}
\def \min{\operatornamewithlimits{min}}
\def \max{\operatornamewithlimits{max}}
\def \log{\operatorname{log}}
\def \max{\operatorname{max}}
\def \rank{\operatorname{rank}}
\def \out{\operatorname{out}}
\def \exp{\operatorname{exp}}
\def \arg{\operatorname{arg}}
\def \E{\operatorname{E}}
\def \tr{\operatorname{tr}}
\def \SNR{\operatorname{SNR}}
\def \dB{\operatorname{dB}}
\def \ln{\operatorname{ln}}

\def \bmat{ \begin{bmatrix} }
\def \emat{ \end{bmatrix} }

\def \be {\begin{eqnarray}}
\def \ee {\end{eqnarray}}
\def \ben {\begin{eqnarray*}}
\def \een {\end{eqnarray*}}

\begin{abstract}
The cost-effective impact and fundamental limits of infrastructure
support with {\it rate-limited} wired backhaul links (i.e., {\em
GreenInfra} support), directly connecting base stations (BSs), are
analyzed in a large-scale hybrid network of unit node density,
where multi-antenna BSs are deployed. We consider a general
scenario such that the rate of each BS-to-BS link scales at an
arbitrary rate relative to the number of randomly located wireless
nodes, $n$. For the operating regimes with respect to the number
of BSs and the number of antennas at each BS, we first analyze the
minimum rate of each backhaul link, $C_{\textrm{BS}}$, required to
guarantee the same throughput scaling as in the infinite-capacity
backhaul link case. We then identify the operating regimes in
which the required rate $C_{\textrm{BS}}$ scales slower than
$n^\epsilon$ for an arbitrarily small $\epsilon>0$ (i.e., the
regimes where $C_{\textrm{BS}}$ does not need to be infinitely
large). We also show the case where our network with GreenInfra is
fundamentally in the {\it infrastructure-limited} regime, in which
the performance is limited by the rate of backhaul links. In
addition, we derive a generalized throughput scaling law including
the case where the rate of each backhaul link scales slower than
$C_{\textrm{BS}}$. To validate the throughput scaling law for
finite values of system parameters, numerical evaluation is also
shown via computer simulations.
\end{abstract}


\newpage

\section{Introduction}\label{SEC:Introduction}
%
%
%
%

In~\cite{GuptaKumar:00}, the sum-rate scaling was originally
introduced and characterized in large wireless {\it ad hoc}
networks. In practice, however, there will be a long latency and
insufficient energy with only wireless connectivity in ad hoc
networks. Hence, it would be good to deploy infrastructure nodes,
or equivalently base stations (BSs), in the network model, i.e.,
hybrid networks, thereby possibly improving the throughput
scaling.

In most network applications, energy efficiency and cost
effectiveness are key performance measures for
greenness~\cite{NiuWuGongYang,HanHarroldArmourKrikidisVidevGrantHaasThompsonKuWangLeNakhaiZhangHanzo,OhKrishnamachariLiuNiu,AshrafBoccardiHo,FettweisZimmermann:08,ShinYiTarokh}
along with the throughput. This is because 1) the operating cost
of BSs, driven in part by inefficient diesel power generators, is
one of the largest concerns and 2) given the economic and
environmental cost, there is a need of environmental grounds to
reduce the energy requirement of network architecture. From a
telecommunications operator's point of view, the following two
benefits can arise from green communications: reduced energy and
network deployment costs, and improved environmental effects.
Especially for large-scale hybrid networks with long-distance
wired backhaul links connecting all the BSs, the {\em deployment
cost of infrastructure} is a critical issue for service providers
since it not only directly relates to the CapEx (capital
expenditures) as well as OpEx (operational
expenditures)~\cite{ChenZhangXu:10,ChenZhangXuYeLi:11} but also
can have significant environmental
impacts~\cite{FehskeFettweisMalmodinBiczok:11,MancusoAlouf:11,DespinsLabeauNgocLabelleCherietThibeaultGagnonGarciaCherkaouiArnaudArnaudMcNeillLemieuxLemay:11}.
Moreover, the increasing number of small cells in the fifth
generation (5G) will result in a significant growth in the number
of connections between cell sites. First of all, high CapEx/OpEx
associated with high backhaul costs may limit the usefulness of
such infrastructure-supported protocols in large hybrid
networks~\cite{ZemlianovVeciana:05,O.Dousse:INFOCOM02,KulkarniViswanath:03,KozatTassiulas:03,LiuLiuTowsley:03,LiuThiranTowsley:07,ShinJeonDevroyeVuChungLeeTarokh:08}.
It is thus vital to significantly dimension the backhaul bandwidth
(or equivalently the backhaul capacity) to reduce the cost of the
operators~\cite{TipmongkolsilpZaghloulJukan:CST11}.

\subsection{Main Contributions}

In this paper, by taking into account greenness in hybrid
networks, we introduce a more general hybrid network with {\em
cost-effective} infrastructure (named {\em GreenInfra}), where the
rate of each BS-to-BS link scales at an arbitrary rate relative to
the number of nodes, $n$. GreenInfra nodes equipped with a large
number of antennas are deployed in the network, and the best among
the two BS-supported schemes
in~\cite{ShinJeonDevroyeVuChungLeeTarokh:08}, i.e.,
infrastructure-supported single-hop (ISH) and
infrastructure-supported multihop (IMH) protocols, is used to
characterize an aggregate throughput of the network. More
precisely, only infrastructure-supported routing protocols are
used for analysis since our main focus is on how the rate of each
backhaul link needs to effectively scale. That is, no pure ad hoc
routing protocols such as multihop (MH)~\cite{GuptaKumar:00} and
hierarchical cooperation~\cite{OzgurLevequeTse:07} are taken into
account. Then, we generalize the throughput scaling result
achieved by the ISH and IMH protocols with an arbitrary scaling of
each backhaul link, which is not definitely straightforward.

Our results present a cost-effective approach for the deployment
of backhaul links connecting a great deal of cell sites (i.e.,
BSs). In order to provide a cost-effective solution to the design
of GreenInfra, we first derive the minimum rate of each BS-to-BS
link, denoted by $C_{\textrm{BS}}$, required to guarantee the same
throughput scaling as in the network using infinite-capacity
backhaul links. The required backhaul link rate $C_{\textrm{BS}}$
is shown according to the two-dimensional operating regimes with
respect to the number of BSs and the number of antennas at each
BS. This backhaul link rate is determined by the multiplication of
the number of matched source--destination (S--D) pairs between any
different two cells and the BS-to-BS transmission rate for each
S--D pair, which is not straightforward since the number of active
S--D pairs between two cells and the BS-to-BS transmission rate
vary according to the operating regimes. Surprisingly, it turns
out that, for some operating regimes, the required backhaul link
rate $C_{\textrm{BS}}$ can be indeed sufficiently small, which
scales much slower than $n^\epsilon$ for an arbitrarily small
$\epsilon>0$, and thus does not need to be infinitely large.

Large-scale ad hoc networks are shown to be fundamentally
power-limited and/or
bandwidth-limited~\cite{OzgurJohariTseLeveque:10}. In addition, we
are interested in further identifying the {\em
infrastructure-limited} regime in which the routing protocol
cannot achieve its maximum throughput scaling due to the small
backhaul link rate. From the infrastructure-limited regime, one
can find the operating regimes, in which the associated throughput
scaling can be improved by increasing the backhaul link rate for
network design. Finally, we analyze the aggregate throughput
scaling for realistic hybrid networks including the case where the
rate of each backhaul link scales slower than $C_{\textrm{BS}}$,
which is based on the derivation of the transmission rate for each
infrastructure-supported routing protocol. To validate the
throughput scaling law for finite values of system parameters, we
also provide numerical results via comprehensive computer
simulations, which are shown to be consistent with our
achievability results.

Our results indicate that a judicious rate scaling of each
BS-to-BS link under a given operating regime leads to the order
optimality of our general hybrid network along with GreenInfra. In
other words, we can still achieve the optimal throughput scaling
of the network by significantly reducing the backhaul link rate
(equivalently, the cost of backhaul links). On the other hand, we
note that, in the next generation communications, the wireless
backhaul is considered as an alternative to wired
backhaul~\cite{TipmongkolsilpZaghloulJukan:CST11}. Our generalized
scaling result provides the theoretical limit on the performance
of the wireless backhaul whose link rate can be smaller than the
minimum required rate.

Our main contribution is fourfold as follows:
\begin{itemize}
  \item As a cost-effective backhaul solution, we derive the minimum rate of each backhaul link required to achieve the same throughput scaling law as in the infinite-capacity backhaul link case.
  \item To better understand the fundamental capabilities of our hybrid network with GreenInfra, we explicitly identify the infrastructure-limited regime according to the number of BSs, the number of antennas at each BS, and the rate of each backhaul link.
  \item To show a more general achievability result, we derive the aggregate throughput scaling with respect to an arbitrary rate scaling of each backhaul link.
  \item We show the numerical results via computer
  simulations.

\end{itemize}

\subsection{Organization}
The rest of this paper is organized as follows. In
Section~\ref{SEC:System}, system and channel models are described.
Routing protocols with infrastructure support are presented in
Section~\ref{SEC:Review}. In Section~\ref{SEC:Routing}, the
minimum required rate of each BS-to-BS link to achieve the
capacity scaling with infinite-capacity backhaul is derived. The
infrastructure-limited regime is identified in
Section~\ref{SEC:InfraLimitedRegime}. The general throughput
scaling with an arbitrary rate scaling of each BS-to-BS link is
also analyzed in Section~\ref{SEC:GeneralizedScaling}. In
Section~\ref{SEC:NumericalResults}, the numerical results are
presented. Finally, we summarize our paper with some concluding
remarks in Section~\ref{SEC:Conclusion}.

\subsection{Notations}
Bold upper and lower case letters denote matrices and vectors,
respectively. The superscript $T$ and $\dagger$ denote the
transpose and conjugate transpose, respectively, of a matrix (or a
vector). The $N\times N$ identity matrix is denoted by
$\mathbf{I}_N$. The expectation is denoted by $\mathbb{E}[\cdot]$.

\section{Previous Work}
It was shown that, for the dense network having $n$ nodes,
randomly distributed in a unit area, the total throughput scales
as $\Theta(\sqrt{n/\log n})$~\cite{GuptaKumar:00}.\footnote{We use
the following notation: i) $f(x)=O(g(x))$ means that there exist
constants $C$ and $c$ such that $f(x)\leq Cg(x)$ for all $x>c$,
ii) $f(x)=o(g(x))$ means that $\lim_{x\rightarrow
\infty}\frac{f(x)}{g(x)}=0$, iii) $f(x)=\Omega(g(x))$ if
$g(x)=O(f(x))$, iv) $f(x)=w(g(x))$ if $g(x)=o(f(x))$, v)
$f(x)=\Theta(g(x))$ if $f(x)=O(g(x))$ and $g(x)=O(f(x))$.} This
throughput scaling is achieved by the MH scheme, where packets of
a source are conveyed to the corresponding destination using the
nearest-neighbor MH transmission. There have been further studies
on MH in the
literature~\cite{GuptaKumar:03,XueXieKumar:05,ShinChungLee:TIT13},
while the total throughput scales far less than $\Theta(n)$.
In~\cite{OzgurLevequeTse:07}, the throughput scaling of the
network having unit area was improved to an almost linear scaling,
i.e., $\Theta(n^{1-\epsilon})$ for an arbitrarily small
$\epsilon>0$, by using a hierarchical cooperation strategy, where
packets of a source are delivered to the corresponding destination
using a long-range multiple-input multiple-output transmission
between clusters recursively. Besides the hierarchical cooperation
scheme~\cite{OzgurLevequeTse:07,NiesenGuptaShah:09}, there have
been various research directions to improve the dense network
throughput up to a linear scaling by using node
mobility~\cite{GrossglauserTse:02}, interference
alignment~\cite{CadambeJafar:08}, directional
antennas~\cite{LiZhangFang:TMC11}, and infrastructure
support~\cite{ZemlianovVeciana:05}.



To further improve the throughput performance with low latency and
low energy, hybrid networks consisting of both wireless ad hoc
nodes and infrastructure nodes have been extensively studied in
\cite{ZemlianovVeciana:05,O.Dousse:INFOCOM02,KulkarniViswanath:03,KozatTassiulas:03,LiuLiuTowsley:03,LiuThiranTowsley:07,ShinJeonDevroyeVuChungLeeTarokh:08,WangLiJiangTangLiu:TMC11}
by showing that BSs can be indeed beneficial in improving the
network throughput. Moreover, one of the most viable ways to meet
growing traffic demands for the high data rate is to use
large-scale (massive) multiple antennas at each
BS~\cite{GuthyUtschickHonig:JSAC13,YangMarzetta:JSAC13}---such
large-scale multiple antenna systems can be thought to be easily
implemented in very high frequency bands (e.g., millimeter wave
bands~\cite{PiKhan:CM11}). Especially, in a hybrid network where
each BS is equipped with a large number of antennas, the optimal
capacity scaling was characterized
in~\cite{ShinJeonDevroyeVuChungLeeTarokh:08}---the achievability
result is based on using one of two infrastructure-supported
routing protocols, i.e, ISH and IMH protocols, pure MH
transmission, and hierarchical cooperation strategy.

It is hardly realistic to assume that BSs are interconnected by
infinite-capacity wired links in hybrid
networks~\cite{ZemlianovVeciana:05,O.Dousse:INFOCOM02,KulkarniViswanath:03,KozatTassiulas:03,LiuLiuTowsley:03,LiuThiranTowsley:07,ShinJeonDevroyeVuChungLeeTarokh:08}.
Hence, it is fundamentally important to characterize a new hybrid
network with {\em rate-limited} backhaul links.
In~\cite{A.Sanderovich:TIT09,O.Simeone:11}, finite-capacity
backhaul links between BSs were taken into account in studying
performance of the multi-cell processing in cooperative cellular
systems based on Wyner-type models which simplify practical
cellular systems. In~\cite{C.Capar:11,C.Capar:12}, the throughput
scaling laws were studied for one- and two-dimensional hybrid
networks, where the wired link interconnecting BSs is
rate-limited. However, the network model under consideration was
comparatively simplified, and the form of achievable schemes was
limited only to MH routings.




\section{System and Channel Models} \label{SEC:System}

\begin{figure}[t!]
  \centering
  \leavevmode \epsfxsize=4.1in   
  \epsffile{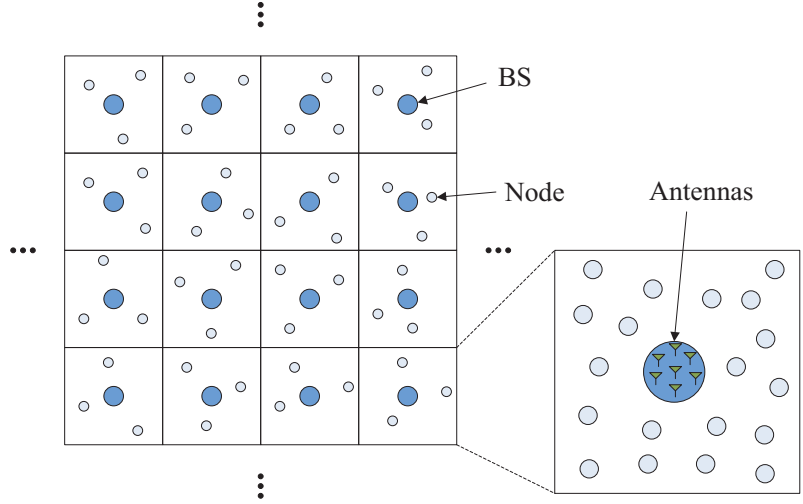}
  \caption{The hybrid network with infrastructure nodes.}
  \label{Fig:Network}
\end{figure}

We consider an extended network of unit node density, where $n$
nodes are uniformly and independently distributed on a square of
area $n$, except for the area covered by BSs. It is assumed that a
source and its destination are paired randomly, so that each node
acts as a source and has exactly one corresponding destination
node. Assume that the BSs are neither sources nor destinations. As
illustrated in Fig.~\ref{Fig:Network}, the whole area of the
network is divided into $m$ square cells of equal area. At the
center of each cell, there is one BS equipped with $l$ antennas.
The total number of antennas in the network is assumed to scale at
most linearly with $n$, i.e., $ml=O(n)$. This network
configuration basically follows that
of~\cite{ShinJeonDevroyeVuChungLeeTarokh:08}.

For analytical convenience, the parameters $n$, $m$, and $l$ are
related according to $n=m^{1/\beta}=l^{1/\gamma}$, where
$\beta,\gamma\in[0,1)$ with a constraint $\beta+\gamma \leq 1$. It
is assumed that BSs are directly interconnected by wired
links.\footnote{In practice, packets are not directly delivered
from a BS to another BS. The packets arrived at a certain BS in a
radio access network are delivered to a core network, and then are
transmitted from the core network to other BSs in the same (or
another) radio access network. For analytical tractability, we
assume direct BS-to-BS communications via backhaul links and then
analyze fundamental limits of the rather simple
infrastructure-supported network model, similarly as in the
previous
work~\cite{ZemlianovVeciana:05,O.Dousse:INFOCOM02,KulkarniViswanath:03,KozatTassiulas:03,LiuLiuTowsley:03,LiuThiranTowsley:07,ShinJeonDevroyeVuChungLeeTarokh:08}.}
In the previous work
\cite{ZemlianovVeciana:05,O.Dousse:INFOCOM02,KulkarniViswanath:03,KozatTassiulas:03,LiuLiuTowsley:03,LiuThiranTowsley:07,ShinJeonDevroyeVuChungLeeTarokh:08},
it is assumed that the rate of BS-to-BS links is unlimited so that
the links are not a bottleneck when packets are delivered from one
cell to another. In practice, however, each BS-to-BS link has a
finite capacity that may limit the transmission rate of
infrastructure-supported routing protocols. In this paper, it is
assumed that each BS is connected to each other through an
errorless wired link with {\em finite rate}
$R_{\textrm{BS}}=n^{\eta}$ for $-\infty<\eta<\infty$.


The uplink channel vector between node $i$ and BS $s$ is denoted by 
\begin{align} \label{EQ:uplinkCH}
    \mathbf{h}_{si}^{(u)}=\left[\frac{e^{j\theta_{si,1}^{(u)}}}{r_{si,1}^{\alpha/2}},
\frac{e^{j\theta_{si,2}^{(u)}}}{r_{si,2}^{\alpha/2}},\ldots,
\frac{e^{j\theta_{si,l}^{(u)}}}{r_{si,l}^{\alpha/2}}\right]^T,
\end{align}
where $\theta_{si,t}^{(u)}$ represents the random phases uniformly
distributed over $[0,2\pi)$ based on a far-field assumption, which
is valid if the wavelength is sufficiently
small~\cite{ShinJeonDevroyeVuChungLeeTarokh:08}. Here, $r_{si,t}$
denotes the distance between node $i$ and the $t$th antenna of BS
$s$, and $\alpha>2$ denotes the path-loss exponent. The downlink
channel vector between BS $s$ and node $i$ is similarly denoted by
$\mathbf{h}_{is}^{(d)}=\left[\frac{e^{j\theta_{is,1}^{(d)}}}{r_{si,1}^{\alpha/2}},
\frac{e^{j\theta_{is,2}^{(d)}}}{r_{si,2}^{\alpha/2}},\ldots,
\frac{e^{j\theta_{is,l}^{(d)}}}{r_{si,l}^{\alpha/2}}\right]$.
The channel between nodes $i$ and $k$ is given by
$h_{ki}=\frac{e^{j\theta_{ki}}}{r_{ki}^{\alpha/2}}$. For the
uplink-downlink balance, we assume that each BS satisfies an
average transmit power constraint $nP/m$, while each node
satisfies an average transmit power constraint $P$. Hence, the
total transmit power of all BSs is the same as the total transmit
power consumed by all wireless nodes. This assumption on the
transmit power is based on the same argument as duality connection
between multiple access channel and broadcast
channel~\cite{ViswanathTse:03}.

It is assumed that the radius of each BS scales as
$\epsilon_0\sqrt{n/m}$, where $\epsilon_0>0$ is an arbitrarily
small constant independent of $n$, $m$, and $l$. This radius
scaling would ensure enough separation among the antennas provided
that per-antenna distance scales (at least) as the average
per-node distance $\Omega(1)$ for any parameters $n$, $m$, and
$l$. If the radius scaling scales slower than $\Theta(1)$, then
per-antenna distance may become vanishingly small, which is
undesirable under our infrastructure-supported routing protocols.
This antenna configuration basically follows the previous
framework established
in~\cite{ShinJeonDevroyeVuChungLeeTarokh:08,GomezRanganErkip:ISIT14}.
Along with the radius scaling of each BS, the antennas of each BS
are placed as follows:\footnote{\textcolor{black}{This antenna
placement strategy guarantees both the nearest-neighbor
transmission from/to each antenna on the BS boundary and the
enough spacing between the antennas of each BS, thus enabling our
IMH protocol to operate properly. If we assume a uniform placement
of antennas inside the BS boundary, then the transmission rate may
be reduced due to a relatively long hop distance between an
antenna and the nearest-neighbor node. Hence, it is natural to
place BS antennas first on the BS boundary. It is worth noting
that the routing protocols, which will be specified in the next
section, can achieve the optimal throughput scaling law under this
antenna configuration~\cite{ShinJeonDevroyeVuChungLeeTarokh:08}.}}
\begin{enumerate}
  \item If $l=w(\sqrt{n/m})$ and $l=O(n/m)$, then $\sqrt{n/m}$ antennas are regularly placed on the BS boundary, i.e., the outermost circle of the BS area, and the remaining antennas are uniformly placed inside the boundary.
  \item If $l=O(\sqrt{n/m})$, then $l$ antennas are regularly placed on the BS boundary.
\end{enumerate}
This antenna scaling can be taken into account in large-scale
(massive) multiple antenna
systems~\cite{GuthyUtschickHonig:JSAC13,YangMarzetta:JSAC13}.

The per-node throughput of the network $R_n$ is assumed to be the
average transmission rate measured in bits or packets per unit
time. Then, the aggregate throughput of the network is defined as
$T_n=nR_n$. The scaling exponent of the aggregate throughput is
defined as\footnote{To simplify notations,
$e(\alpha,\beta,\gamma,\eta)$ will be written as $e$ if dropping
$\alpha$, $\beta$, $\gamma$, and $\eta$ does not cause any
confusion.}
\begin{align}
e(\alpha,\beta,\gamma,\eta)=\lim_{n\rightarrow\infty}\frac{\log
T_n(\alpha,\beta,\gamma,\eta)}{\log
    n}. \nonumber
\end{align}
We will later examine the scaling exponent
$e(\alpha,\beta,\gamma,\eta)$ for routing protocols in an
operating regime that is identified according to the path-loss
exponent, the number of BSs, the number of antennas per BS, and
the backhaul link rate.



\section{Hybrid Network With Infinite-Capacity Infrastructure} \label{SEC:Review}

In this section, the overview of routing protocols with
infrastructure support and the throughput scaling results under
the protocols are provided for readability of remaining sections.


\subsection{Routing Protocols With Infrastructure Support}

The order-optimal routing protocols supported by BSs having
multiple antennas in~\cite{ShinJeonDevroyeVuChungLeeTarokh:08} are
described in the following. With the aid of the
infrastructure-supported routing protocols, packets of a source
are delivered to the corresponding destination of the source using
three stages: {\it access routing}, {\it BS-to-BS communication},
and {\it exit routing}.
According to the transmission scheme in the access and exit
routings, the infrastructure-supported routing protocols are
categorized into two different protocols as in the following.

\begin{figure}[t!]
  \centering
  \leavevmode \epsfxsize=4.5in
  \epsffile{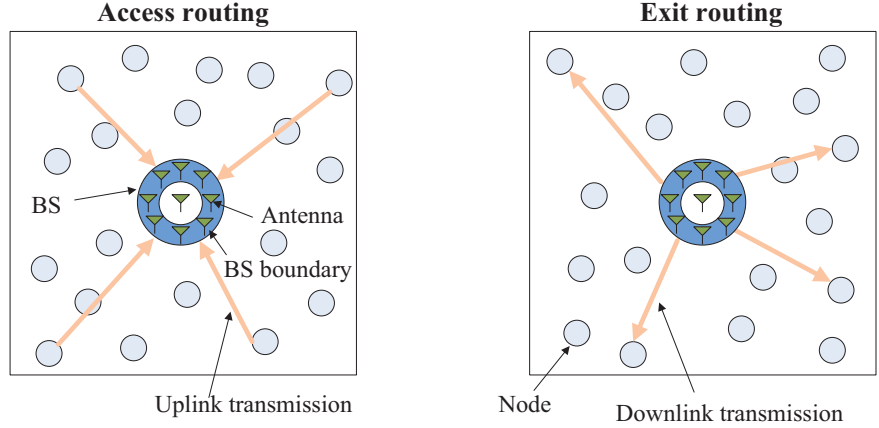}
  \caption{The ISH protocol. Each square represents a cell in the hybrid network.}
  \label{Fig:ISHprotocol}
\end{figure}

\subsubsection{ISH Protocol} \label{SEC:ISH}
In the ISH protocol, packets of source nodes are transmitted to
the BSs via single-hop multiple access, and the BSs transmit the
packets to the destination nodes via single-hop broadcast (see
Fig.~\ref{Fig:ISHprotocol}, in which two cells are shown). The ISH
protocol is described as follows.
\begin{itemize}
  \item There are $n/m$ nodes with high probability (whp) in each cell~\cite[Lemma 4.1]{OzgurLevequeTse:07}.
  \item For the access routing, all source nodes in each cell transmit their packets simultaneously to the home-cell BS via single-hop multiple-access.
  \item The packets of source nodes are then jointly decoded at each BS, assuming that the signals transmitted from the other cells are treated as noise. The minimum mean-square error (MMSE) estimation with successive interference cancellation (SIC) is performed at each BS. More precisely, the $l\times 1$ receive filter $\mathbf{v}_i$ for the signal of the $i$-th node at BS $s$ is given by~\cite{ViswanathTse:03}
      \begin{align}
      \mathbf{v}_i =
      \left(\mathbf{I}_l+\sum_{k>i}P\mathbf{h}_{sk}^{(u)}\mathbf{h}_{sk}^{(u)\dagger}\right)^{-1}\mathbf{h}_{si}^{(u)},\label{EQ:MMSE}
      \end{align}
      where the signals from nodes $1,\cdots,i-1$ are cancelled and the signals from nodes $i+1,\cdots,n/m$ are treated as noise when the cancelling order is $1,\cdots,n/m$.
  \item In the next stage, the decoded packets are transmitted to the BS nearest to the corresponding destination of the source via wired BS-to-BS link.
  \item For the exit routing, the BS in each cell transmits $n/m$ packets received from other cells to the wireless nodes in its cell via single-hop broadcast. The transmit precoding in the downlink is designed by the dual system of the receive filters in the uplink. The $l\times 1$ transmit precoding vector $\mathbf{u}_i$ with dirty paper coding (DPC) at BS $s$ is given by~\cite{Costa:83}
      \begin{align}
      \mathbf{u}_i =
      \left(\mathbf{I}_l+\sum_{k>i}p_k\mathbf{h}_{ks}^{(d)\dagger}\mathbf{h}_{ks}^{(d)}\right)^{-1}\mathbf{h}_{is}^{(d)\dagger},
      \label{EQ:DPC}
      \end{align}
      where the power $p_k\geq 0$ is allocated to each node such that $\sum_k p_k \leq \frac{nP}{m}$ for $k=1,\cdots,n/m$.
\end{itemize}

\subsubsection{IMH Protocol} \label{SEC:IMH}
\begin{figure}[t!]
  \centering
  \leavevmode \epsfxsize=4.5in
  \epsffile{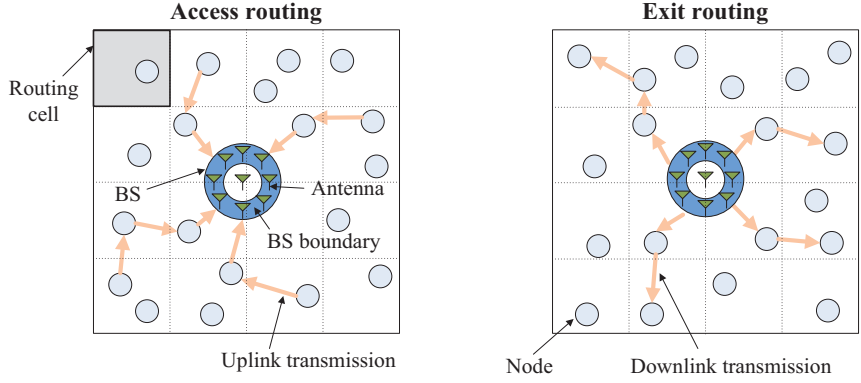}
  \caption{The IMH protocol. Each square represents a cell in the hybrid network. The shaded square in the left figure represents a routing cell.}
  \label{Fig:IMHprotocol}
\end{figure}

Since the extended network is fundamentally
power-limited~\cite{OzgurLevequeTse:07,OzgurJohariTseLeveque:10},
the ISH protocol may not be effective especially when the node-BS
distance is quite long. Thus, we introduce the IMH protocol (see
Fig. \ref{Fig:IMHprotocol}, in which two cells are shown). In the
IMH protocol, the packets are transmitted between nodes and BSs
using MH routing.
\begin{itemize}
  \item Each cell is further divided into smaller square cells of area $2\log n$, termed routing cells.
  \item When $\min\{l,\sqrt{n/m}\}$ antennas are regularly placed on the BS boundary, $\min\{l,\sqrt{n/m}\}$ MH paths can be used simultaneously.\footnote{We use antennas only on the BS boundary for the IMH protocol since it may cause a performance degradation to use MH transmission between the antennas inside the boundary and the nearest-neighbor nodes due to a relatively longer hop distance. In practice, these unused antennas may not be deployed if the IMH protocol is used only.}
  \item For the access routing, the antennas placed only on the home-cell BS boundary can receive the packet transmitted from one of the nodes in the nearest-neighbor routing cell. The MH routing is performed horizontally or vertically by using the adjacent routing cells passing through the line connecting a source to one of the antennas of its BS. The transmission rate scaling per routing path does not depend on
the path loss exponent $\alpha$ since the
signal-to-interference-and-noise ratio (SINR) seen by any receiver
(a BS antenna) is given by $\Omega(1)$ owing to the
nearest-neighbor routing.
  \item The BS-to-BS communication is the same as the ISH protocol.
  \item For the exit routing, each antenna on the target-cell BS boundary transmits the packets to one of the nodes in the nearest-neighbor routing cell. The packets are transmitted along a line connecting the antenna of its BS to the corresponding
  destination. The SINR at the receiving node is
  also given by $\Omega(1)$.
\end{itemize}

\subsection{The Throughput Scaling With Infinite Backhaul Link Rate}
In this subsection, the throughput scaling results of the two
routing protocols with infinite backhaul link rate derived in
\cite{ShinJeonDevroyeVuChungLeeTarokh:08} are summarized.

\begin{lemma}[\cite{ShinJeonDevroyeVuChungLeeTarokh:08}]\label{Lem:RateISH}
Suppose that the ISH protocol is used in the extended network,
where the rate of each BS-to-BS link is unlimited. Then, the
aggregate rate scaling is given by
\begin{align}\label{Eq:ThroughputISH-BSunlimited}
    T_{n,\textrm{ISH}}=\Omega\left( ml\left(\frac{m}{n}\right)^{\alpha/2-1}\right).
\end{align}
\end{lemma}

\begin{lemma}[\cite{ShinJeonDevroyeVuChungLeeTarokh:08}]\label{Lem:RateIMH}
Suppose that the IMH protocol is used in the extended network,
where the rate of each backhaul link is unlimited. Then, the
aggregate rate scaling is given by
\begin{align}\label{Eq:ThroughputIMH-BSunlimited}
    T_{n,\textrm{IMH}}=\Omega\left(m\min\left\{
    l,\left(\frac{n}{m}\right)^{1/2-\epsilon}\right\}\right),
\end{align}
where $\epsilon>0$ is an arbitrarily small constant.
\end{lemma}

Note that unlike the ISH protocol, the aggregate rate scaling of
the IMH protocol does not depend on the path loss exponent
$\alpha$ since the IMH is designed based on the nearest-neighbor
MH. From Lemmas \ref{Lem:RateISH} and \ref{Lem:RateIMH}, the
aggregate throughput of the network is given by the maximum of
these two scaling laws as in the following theorem.

\begin{theorem}\label{Lem:TotalThroughputInfinite}
In the hybrid network of unit node density, where the rate of each
backhaul link is unlimited, the aggregate throughput achieved by
both ISH and IMH protocols is given by
\begin{align}\label{Eq:TotalThroughputInfinite} 
    T_n
    &=
    \max\left\{T_{n,\textrm{ISH}},T_{n,\textrm{IMH}}\right\}
    \nonumber\\
    &=\Omega\left(
    \max\left\{ml\left(\frac{m}{n}\right)^{\alpha/2-1},m\min\left\{l,\left(\frac{n}{m}\right)^{1/2-\epsilon}\right\}
    \right\}\right),
\end{align}
where $\epsilon>0$ is an arbitrarily small constant.
\end{theorem}

In order to better understand the aggregate throughput in
(\ref{Eq:TotalThroughputInfinite}), two operating regimes with
respect to the scaling parameters $\beta$ and $\gamma$ are
identified as in
Fig.~\ref{Fig:OperatingRegimesInfiniteBScapacity}. For each
regime, we can determine the best routing scheme between ISH and
IMH by comparing their throughput scaling exponents. The best
routing scheme and its condition are summarized in
TABLE~\ref{Tab:RateExtended}. The throughput achieved by the ISH
protocol gets improved with increasing number of antennas per BS,
$l$. Thus, the ISH protocol can be used in Regime B. At the high
path-loss attenuation regime, since the network is power-limited,
the IMH protocol becomes dominant.



\ifx \doubleColumn \undefined 
\begin{figure}[t!] 
  \centering
  \leavevmode \epsfxsize=3.2in   
  \epsffile{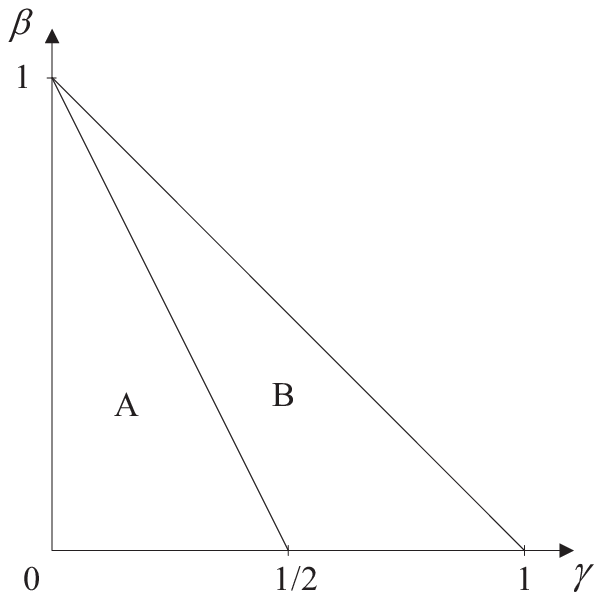}
  \caption{The operating regimes on the throughput scaling with respect to $\beta$ and $\gamma$. Regimes A and B are the set of operating points satisfying $\beta+2\gamma<1$ and $\beta+2\gamma\geq 1$, respectively.}
  \label{Fig:OperatingRegimesInfiniteBScapacity}
\end{figure}
\else
\begin{figure}[t!] 
  \centering
  \leavevmode \epsfxsize=1.8in   
  \epsffile{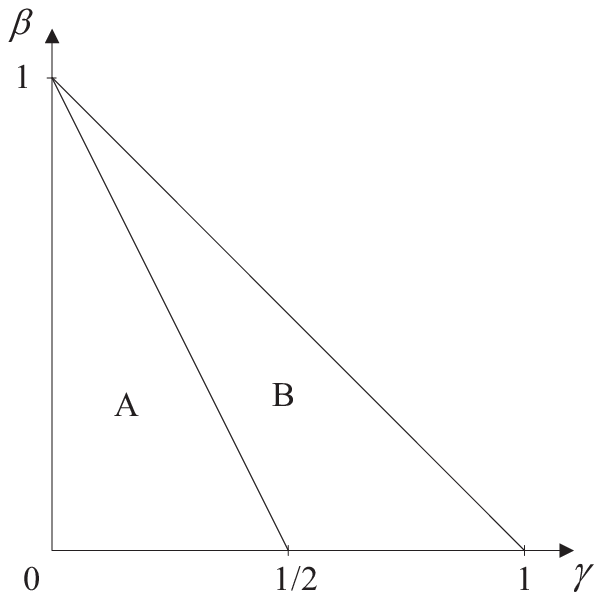}
  \caption{The operating regimes on the throughput scaling with respect to $\beta$ and $\gamma$. Regimes A and B are the set of operating points satisfying $\beta+2\gamma<1$ and $\beta+2\gamma\geq 1$, respectively.}
  \label{Fig:OperatingRegimesInfiniteBScapacity}
\end{figure}
\fi

\begin{table}[t]
    \centering
    \caption{Throughput scaling for an extended network with infinite-capacity infrastructure}
    \label{Tab:RateExtended}
\begin{tabular}{|c|c|c|c|}
  \hline
  Regime & Condition & Scheme & $e(\alpha,\beta,\gamma,\infty)$\\
  \noalign{\hrule height 1.2pt}
  \hline
  \multirow{1}{*}{A} & $\alpha>2$  & IMH & $\beta+\gamma$\\
  \hline
  \multirow{2}{*}{B} & $\alpha<1+\frac{2\gamma}{1-\beta}$  & ISH & $1+\gamma-\frac{\alpha(1-\beta)}{2}$\\
                     & $\alpha\geq 1+\frac{2\gamma}{1-\beta}$ & IMH & $\frac{1+\beta}{2}$\\
  \hline
\end{tabular}
\end{table}


\section{The Design of Cost-Effective Infrastructure} \label{SEC:Routing}

As the number of BSs, $m$, increases, the number of
interconnections between BSs rapidly grows at a rate of
$\Theta(m^2)$, thus resulting in very high CapEx and OpEx in
designing high-capacity backhaul links especially in a large-scale
hybrid network. Hence, it is crucial to minimize the rate of
backhaul links without sacrificing the aggregate throughput of the
network.

In this section, we will derive the minimum rate of each backhaul
link required to achieve the aggregate throughput scaling in
Theorem~\ref{Lem:TotalThroughputInfinite} by analyzing the number
of active S--D pairs between two cells according to the
two-dimensional operating regimes. The minimum required rate of
BS-to-BS links is determined by the multiplication of the number
of matched S--D pairs between any different two cells and the
transmission rate of the infrastructure-supported protocols for
each S-D pair. Packets of these S--D pairs are conveyed through
the backhaul link between their associated two cells. The number
of matched S--D pairs between any different two cells is first
derived in Lemma~\ref{Lem:NumPairs}. Using
Lemma~\ref{Lem:NumPairs} and the throughput scaling results based
on the use of ISH and IMH routing protocols in
(\ref{Eq:ThroughputISH-BSunlimited}) and
(\ref{Eq:ThroughputIMH-BSunlimited}), respectively, the minimum
required rates of backhaul links for the ISH and IMH routing
protocols are then derived in Lemmas~\ref{Lem:CBS-ISH} and
\ref{Lem:CBS-IMH}, respectively. The minimum required rate of each
backhaul link to guarantee the maximum throughput capacity scaling
in Theorem~\ref{Lem:TotalThroughputInfinite} can be derived by
comparing the required rates of backhaul links for the two
infrastructure-supported protocols. Let us start from the
following lemma, which derives the number of matched S--D pairs
between any two different cells.


\begin{lemma}\label{Lem:NumPairs}
Suppose that there are $n^a$ simultaneously transmitting source
nodes in each cell, where $a$ is a positive constant. A source
node in each cell randomly chooses its destination node that is
placed in one cell among $n^b$ cells, where $b$ is a positive
constant.
Then, the number of destinations in the $k$th cell whose source
nodes are in the $i$th cell, $X_{ki}$, is given by\footnote{Note
that the parameter $X_{ki}$ in
Lemmas~\ref{Lem:NumPairs}--\ref{Lem:CBS-IMH} does not depend on
$k$ and $i$ since each source chooses its destination randomly and
independently. However, since we need indices of $k$ and $i$ in
$X_{ki}$ for an easier proof of Lemma~\ref{Lem:CBS-IMH}, we use
the notation $X_{ki}$ in the main text for notational
consistency.}
\begin{align}\label{Eq:NumSDpairs}
    X_{ki}=\left\{\begin{array}{ll}
    O\left(\log n\right) &\textrm{if~} a\leq b\\
    \Theta \left(n^{a-b}\right) &\textrm{if~} a>b
    \end{array}
    \right.
\end{align}
whp as $n$ tends to infinity, where $i,k\in \{1,\ldots,m\}$.
\end{lemma}

\begin{IEEEproof}
Refer to Appendix~\ref{PF:NumPairs}.
\end{IEEEproof}

\begin{figure}[t!]
  \centering
  \leavevmode \epsfxsize=3.6in
  \epsffile{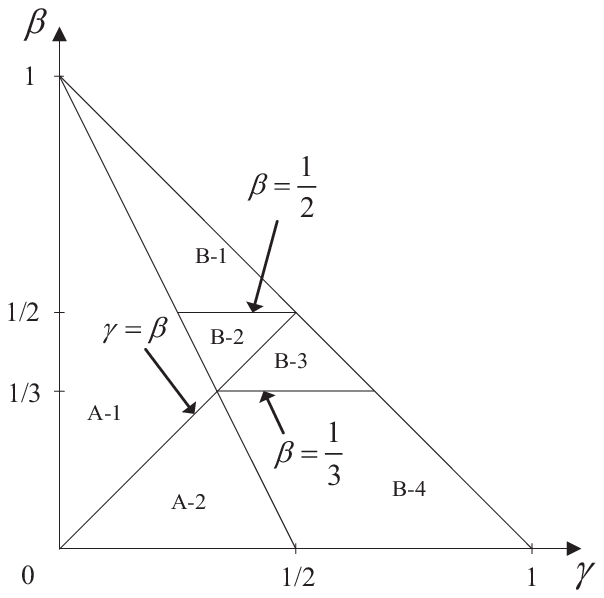}
  \caption{The operating regimes on the required rate of each BS-to-BS link with respect to $\beta$ and $\gamma$.}
  \label{Fig:OperatingRegimesFiniteBScapacity}
\end{figure}

The number of S--D pairs between cells depends on the routing type
as well as two scaling parameters $\beta$ and $\gamma$. For ease
of explanation, we further divide Regimes A and B into smaller
sub-regimes as illustrated in
Fig.~\ref{Fig:OperatingRegimesFiniteBScapacity}.



\subsection{The Minimum Required Rate for the ISH Protocol}

Since the value of $\alpha$ is not manageable but rather affected
by the channel characteristics, the minimum required rate of each
backhaul link should be computed by taking into account the
transmission rate of the ISH protocol maximized over $\alpha$. The
required rate of backhaul links for the ISH protocol is derived in
the following lemma. In this subsection, we focus on Regime B
since the ISH protocol is used only in the regime, as depicted in
TABLE~\ref{Tab:RateExtended}.

\begin{lemma}\label{Lem:CBS-ISH}
Suppose that the ISH protocol is used in Regime B of the network
under consideration. Then, the number of destinations in the $k$th
cell whose source nodes are in the $i$th cell, $X_{ki}$, is given
by
\begin{align}\label{Eq:BoundPacketsISH}
    X_{ki}=\left\{\begin{array}{ll}
    O\left(\log n\right) &\textrm{for Regime B-1}\\
    O\left(n^{1-2\beta}\right) &\textrm{for Regimes B-2, B-3, and B-4,}\\
    \end{array}
    \right.
\end{align}
where $i,k\in \{1,\ldots,m\}$, and $m=n^\beta$. The minimum rate
of each BS-to-BS link required to achieve the throughput in
(\ref{Eq:ThroughputISH-BSunlimited}), $C_{\textrm{BS,ISH}}$, is
given by
\ifx \doubleColumn \undefined 
\begin{align} 
    C_{\textrm{BS,ISH}}
    = \left\{ \begin{array}{ll}
    \Omega \left(n^{\beta+\gamma-1}\log n\right)
    &\textrm{for Regime B-1} \\
    \Omega \left(n^{\gamma-\beta}\right)
     &\textrm{for Regimes B-2, B-3, and B-4,}
    \end{array} \right. \nonumber
\end{align}
\else
\begin{align} 
    C_{\textrm{BS,ISH}}
    = \left\{ \begin{array}{ll}
    \Omega \left(n^{\beta+\gamma-1}\right)
    &\textrm{for Regime B-1} \\
    \Omega \left( n^{\gamma-\beta}\right)
     &\textrm{for Regimes B-2,}
     \\&\textrm{B-3, and B-4}
    \end{array} \right. \nonumber
\end{align}
\fi where $l=n^{\gamma}$.
\end{lemma}

\begin{IEEEproof}
Refer to Appendix~\ref{PF:CBS-ISH}.
\end{IEEEproof}

From Lemma~\ref{Lem:CBS-ISH}, it is examined that, as
$\beta<\frac{1}{2}$ (i.e., in Regimes B-3, and B-4), the minimum
required rate of each backhaul link, $C_{\textrm{BS,ISH}}$, under
the ISH protocol grows rapidly with increasing $n$. This is
because more data traffic needs to be delivered through each
backhaul link in the regime.


\subsection{The Minimum Required Rate for the IMH Protocol}

Since the throughput scaling of the IMH routing protocol has a
different form depending on both scaling parameters $\beta$ and
$\gamma$, deriving the required rate of each BS-to-BS link for the
IMH protocol is not as simple as the ISH routing case. In the
following lemma, the required backhaul link rates for the IMH
protocol, termed $ C_{\textrm{BS,IMH}}$, are characterized with
respect to $\beta$ and $\gamma$ by showing three different rates.
In this subsection, we focus on whole operating regimes since the
IMH protocol can be used in any operating regimes (see
TABLE~\ref{Tab:RateExtended}).

\begin{lemma}\label{Lem:CBS-IMH}
Suppose that the IMH protocol is used in all the operating regimes
of the network under consideration. Then, the number of
destinations in the $k$th cell whose source nodes are in the $i$th
cell, $X_{ki}$, is given by
\ifx \doubleColumn \undefined 
\begin{align}\label{Eq:BoundPacketsIMH} 
    X_{ki}=\left\{\begin{array}{ll}
    O\left(\log n\right) &\textrm{for Regimes A-1, B-1, B-2, and B-3}\\
    O\left(n^{\gamma-\beta}\right) &\textrm{for Regime A-2}\\
    O\left(n^{\frac{1-3\beta}{2}-\epsilon}\right) &\textrm{for Regime B-4},
    \end{array}
    \right.
\end{align}
\else
\begin{align}\label{Eq:BoundPacketsIMH} 
    \mathcal{X}=\left\{\begin{array}{ll}
    O\left(\log n\right) &\textrm{for Regimes  A-1, B-1, B-2,}
    \\&\textrm{and B-3}\\
    O\left(n^{\gamma-\beta}\right) &\textrm{for Regime A-2}\\
    O\left(n^{\frac{1-3\beta}{2}-\epsilon}\right) &\textrm{for Regime B-4,}
    \end{array}
    \right.
\end{align}
\fi where $i,k\in \{1,\ldots,m\}$, $m=n^{\beta}$, $l=n^{\gamma}$,
and $\epsilon>0$ is an arbitrarily small constant. The minimum
rate of each BS-to-BS link required to achieve the throughput in
(\ref{Eq:ThroughputIMH-BSunlimited}), $C_{\textrm{BS,IMH}}$, is
given by
\ifx \doubleColumn \undefined 
\begin{align} 
    C_{\textrm{BS,IMH}} = \left\{ \begin{array}{ll}
    \Omega\left(\log n\right) &\textrm{for Regimes A-1, B-1, B-2, and B-3}\\
    \Omega\left(n^{\gamma-\beta}\right) &\textrm{for Regime A-2}\\
    \Omega\left(n^{\frac{1-3\beta}{2}-\epsilon}\right) &\textrm{for Regime B-4}.
    \end{array} \right. \nonumber
\end{align}
\else
\begin{align} 
    C_{\textrm{BS,IMH}} = \left\{ \begin{array}{ll}
    \Omega\left(\log n\right) &\textrm{for Regimes A-1, B-1, B-2,}
    \\&\textrm{and B-3}\\
    \Omega\left(n^{\gamma-\beta}\right) &\textrm{for Regime A-2}\\
    \Omega\left(n^{\frac{1-3\beta}{2}-\epsilon}\right) &\textrm{for Regime B-4}.
    \end{array} \right. \nonumber
\end{align}
\fi
\end{lemma}

\begin{IEEEproof}
Refer to Appendix~\ref{PF:CBS-IMH}.
\end{IEEEproof}

From Lemma~\ref{Lem:CBS-IMH}, one can see that the minimum
required rate of each backhaul link under the IMH protocol,
$C_{\textrm{BS,IMH}}$, is nebligibly small for Regimes A-1, B-1,
B-2, and B-3. In other words, in these sub-regimes, the data
traffic between BSs is well-distributed to a relatively large
number of BSs over the network.

\subsection{The Minimum Required Rate for the Optimal Infrastructure-Supported Routing Protocols}

Based on Lemmas \ref{Lem:CBS-ISH} and \ref{Lem:CBS-IMH}, we are
now ready to establish our first main theorem, which characterizes
the minimum BS-to-BS link rate guaranteeing the theoretically
maximum throughput scaling in
Theorem~\ref{Lem:TotalThroughputInfinite} when the best between
the ISH and IMH protocols is used according to the operating
regimes in our hybrid network with GreenInfra.

\begin{theorem}\label{Thm:RequiredRate}
In the hybrid network with GreenInfra, the minimum rate of each
BS-to-BS link required to achieve the throughput scaling in
Theorem~\ref{Lem:TotalThroughputInfinite}, $C_{\textrm{BS}}$, is
given by
\ifx \doubleColumn \undefined 
\begin{align} 
    C_{\textrm{BS}} = \left\{ \begin{array}{ll}
    \Omega\left(\log n\right) &\textrm{for Regimes A-1, B-1, and B-2}\\
    \Omega\left(n^{\gamma-\beta}\right) &\textrm{for Regimes A-2, B-3, and B-4}
    \end{array} \right. \label{EQ:C_BS}
\end{align}
\else
\begin{align} 
    C_{\textrm{BS}} = \left\{ \begin{array}{ll}
    \Omega\left(\log n\right) &\textrm{for Regimes A-1, B-1, B-2,}
    \\&\textrm{and A-2}\\
    \Omega\left(n^{\gamma-\beta}\right) &\textrm{for Regimes B-3 and B-4}\\
    \end{array} \right. \label{EQ:C_BS}
\end{align}
\fi where $\epsilon>0$ is an arbitrarily small constant. The
associated operating regimes with respect to scaling parameters
$\beta$ and $\gamma$ are illustrated in
Fig.~\ref{Fig:OperatingRegimesFiniteBScapacity}.
\end{theorem}

\begin{IEEEproof}
In Regime A, only the IMH protocol is used among the two
infrastructure-supported protocols. Thus, $C_{\textrm{BS}}$ is the
same as $C_{\textrm{BS,IMH}}$ in these regimes. However, since
either the ISH or IMH protocol can be used according to the value
of $\alpha$ in Regime B, we should compare the required rates of
backhaul links for both protocols using Lemmas \ref{Lem:CBS-ISH}
and \ref{Lem:CBS-IMH}. We recall that for the ISH protocol, the
required backhaul link rate is maximized over $\alpha$ as in
Lemma~\ref{Lem:CBS-ISH}. Then in Regime B-1, $C_{\textrm{BS,ISH}}$
and $C_{\textrm{BS,IMH}}$ are given by $C_{\textrm{BS,ISH}}=\Omega
\left((\log n) n^{\beta+\gamma-1}\right)$ and
$C_{\textrm{BS,IMH}}=\Omega\left(\log n \right)$, respectively.
Since $\beta+\gamma-1\leq 0$, $C_{\textrm{BS,ISH}}$ is smaller
than or equal to $C_{\textrm{BS,IMH}}$ in Regime B-1, and thus it
follows that $C_{\textrm{BS}}=C_{\textrm{BS,IMH}}=\Omega\left(\log
n \right)$ in Regime B-1.

In Regimes B-2 and B-3, $C_{\textrm{BS,ISH}}$ and
$C_{\textrm{BS,IMH}}$ are given by $C_{\textrm{BS,ISH}}=\Omega
\left(n^{\gamma-\beta} \right)$ and
$C_{\textrm{BS,IMH}}=\Omega\left(\log n \right)$, respectively. It
thus follows that $C_{\textrm{BS,ISH}}<C_{\textrm{BS,IMH}}$ in
Regime B-2 because $\gamma\leq \beta$, while
$C_{\textrm{BS,ISH}}>C_{\textrm{BS,IMH}}$ in Regime B-3 because
$\gamma>\beta$. Hence, we have
\begin{align}
     C_{\textrm{BS}} = \left\{ \begin{array}{ll}
    C_{\textrm{BS,IMH}}=\Omega(\log n) &\textrm{for Regime B-2}\\
    C_{\textrm{BS,ISH}}=\Omega \left(n^{\gamma-\beta} \right)&{\textrm{for Regime B-3}}.
    \end{array} \right. \nonumber
\end{align}

In Regime B-4, $C_{\textrm{BS,ISH}}$ and $C_{\textrm{BS,IMH}}$ are
given by $C_{\textrm{BS,ISH}}=\Omega \left( n^{\gamma-\beta}
\right)$ and
$C_{\textrm{BS,IMH}}=\Omega\left(n^{\frac{1-3\beta}{2}-\epsilon}\right)$,
respectively. The difference of the scaling exponents of
$C_{\textrm{BS,ISH}}$ and $C_{\textrm{BS,IMH}}$ is given by
$\frac{\beta+2\gamma-1}{2}+\epsilon$.
Since $\beta+2\gamma>1$ in Regime B-4, we have
$C_{\textrm{BS,ISH}}> C_{\textrm{BS,IMH}}$ and
$C_{\textrm{BS}}=C_{\textrm{BS,ISH}}=\Omega\left(n^{\gamma-\beta}\right)$
in Regime B-4. Therefore, the minimum required rate of each
BS-to-BS link is finally given by (\ref{EQ:C_BS}), which completes
the proof of Theorem~\ref{Thm:RequiredRate}.
\end{IEEEproof}

In Regimes A-2, B-3, and B-4, the required rate of each backhaul
link is given by $\Omega\left(n^{\gamma-\beta}\right)$. In the
regime, the backhaul link rate $C_{\textrm{BS}}$ is increased when
the number of antennas per BS, $l$, increases, whereas it is
decreased when the number of BSs, $m$, increases.

In Regime B, either the ISH or IMH protocol can be used to achieve
the throughput scaling in
Theorem~\ref{Lem:TotalThroughputInfinite} according to $\alpha$.
Specifically, when $\alpha$ is moderately small, the transmission
rate of the ISH protocol is greater than that of the IMH protocol
as shown in Lemma~\ref{Lem:RateISH}, while the throughput of the
IMH protocol does not depend on $\alpha$. Hence, the required rate
$C_{\textrm{BS}}$ of each backhaul link in Regime B should be
determined by the maximum of the required rates of backhaul links
for the ISH and IMH protocols.

As addressed earlier, the operating regimes A and B are further
divided into smaller sub-regimes according to the minimum required
rates of backhaul links. More specifically, Regime A is divided
into Regimes A-1 and A-2 by the borderline $\gamma=\beta$ since
the number of matched S--D pairs is different for the two cases
$\gamma\leq\beta$ and $\gamma>\beta$ from
Lemma~\ref{Lem:NumPairs}. The division of Regime B can be
explained in a similar fashion.

\begin{remark}[Case of Negligibly Small Backhaul Link Rates]\label{Rem:SmallBackhaulRate}
It is worth noting that, in Regimes A-1, B-1, and B-2, the
required rate of each BS-to-BS link is negligibly small, i.e.,
$C_{\textrm{BS}}=O(n^{\epsilon})$ for an arbitrarily small
$\epsilon>0$. This indicates that the backhaul link rate does not
need to be infinitely high in these sub-regimes even for a large
number of wireless nodes in the network.
\end{remark}

\section{Identification of the Infrastructure-Limited Regime}\label{SEC:InfraLimitedRegime}

The extended network is fundamentally power-limited, i.e., the
aggregate throughput is determined by the power transfer between
S--D pairs~\cite{OzgurLevequeTse:07,OzgurJohariTseLeveque:10}. In
our extended network with GreenInfra, the operating regime can
also be infrastructure-limited, where the backhaul link rate
$R_{\textrm{BS}}$ scales slower than the minimum required rate
$C_{\textrm{BS}}$.


In the following, we explicitly identify the
infrastructure-limited regime depending on $\eta$ in
Figs.~\ref{Fig:InfraLimitedRegimes-2}--\ref{Fig:InfraLimitedRegimes-3}.
From Theorem~\ref{Thm:RequiredRate}, it is seen that, when
$\eta\leq 0$, the rate of each backhaul link, $R_{\textrm{BS}}$,
scales slower than the minimum required rate $C_{\textrm{BS}}$ in
all the operating regimes. Hence, all these regimes are
infrastructure-limited if $\eta\leq 0$. It indicates that, unless
$\eta>0$, one can achieve only a lower throughput scaling than the
ideal throughput scaling in
Theorem~\ref{Lem:TotalThroughputInfinite} for any scaling
parameters $\beta$ and $\gamma$. If the value of $\eta$ is greater
than zero, however, Regimes A-1, B-1, and B-2 becomes not
infrastructure-limited since the required rate of each backhaul
link in these regimes scales much slower than $n^\epsilon$ for an
arbitrarily small $\epsilon>0$, as mentioned in
Remark~\ref{Rem:SmallBackhaulRate}. When $0<\eta<\frac{1}{2}$,
some parts of Regimes A-2, B-3, and B-4 are not
infrastructure-limited, and the area of these parts grows as
$\eta$ increases. Specifically, if $\eta<\gamma-\beta$, then
Regimes A-2, B-3, and B-4 are infrastructure-limited since
$C_{\textrm{BS}}=\Omega(n^{\gamma-\beta})$ from
Theorem~\ref{Thm:RequiredRate}, and the associated part is shaded
in Fig.~\ref{Fig:InfraLimitedRegimes-2}. If
$\frac{1}{2}\leq\eta<1$, then Regimes A-2 and B-3 are no longer
infrastructure-limited but a part of Regime B-4 is still
infrastructure-limited, which is illustrated in
Fig.~\ref{Fig:InfraLimitedRegimes-3}. All operating regimes are
not infrastructure-limited if $\eta\geq 1$.

\ifx \doubleColumn \undefined 
\begin{figure}[t!] 
  \centering
  \leavevmode \epsfxsize=4.7in
  \epsffile{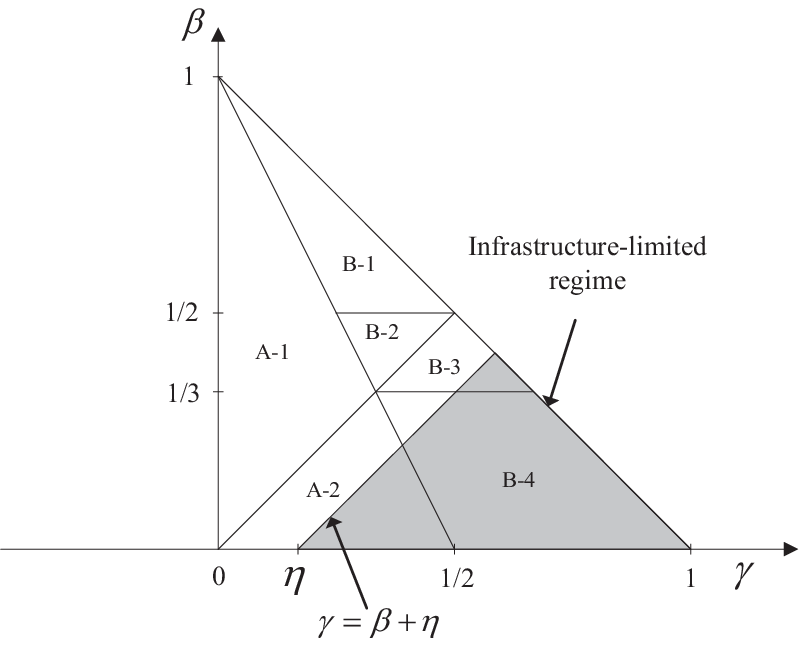}
  \caption{The infrastructure-limited regime with respect to $\beta$ and $\gamma$, where $0<\eta <\frac{1}{2}$.}
  \label{Fig:InfraLimitedRegimes-2}
\end{figure}
\else
\begin{figure}[t!] 
  \centering
  \leavevmode \epsfxsize=3.0in
  \epsffile{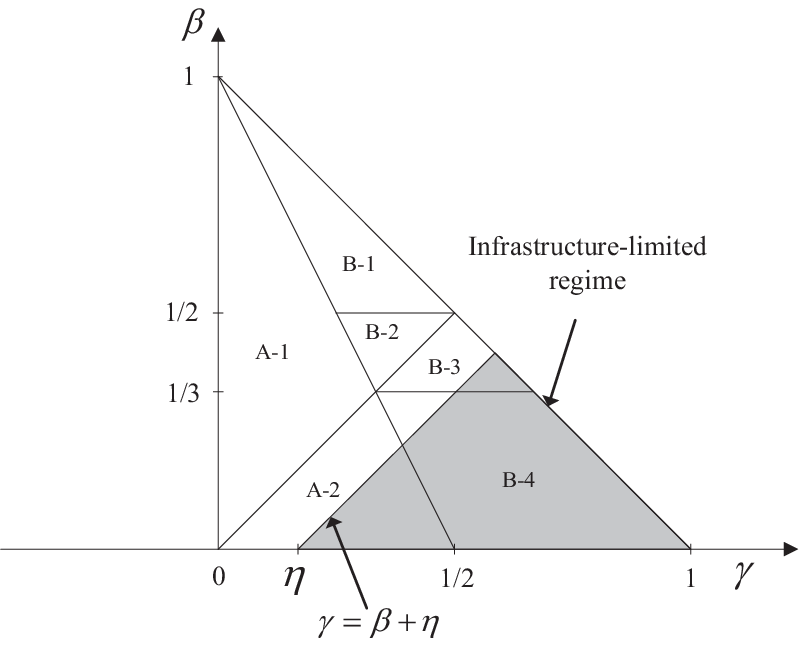}
  \caption{The infrastructure-limited regime with respect to $\beta$ and $\gamma$, where $0<\eta <\frac{1}{2}$.}
  \label{Fig:InfraLimitedRegimes-2}
\end{figure}
\fi

\ifx \doubleColumn \undefined 
\begin{figure}[t!] 
  \centering
  \leavevmode \epsfxsize=4.7in
  \epsffile{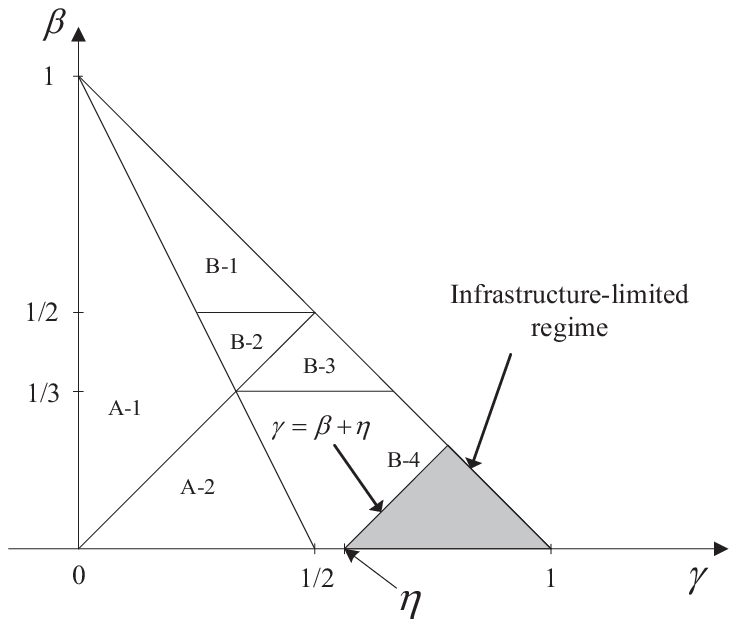}
  \caption{The infrastructure-limited regime with respect to $\beta$ and $\gamma$, where $\frac{1}{2}\leq\eta <1$.}
  \label{Fig:InfraLimitedRegimes-3}
\end{figure}
\else
\begin{figure}[t!] 
  \centering
  \leavevmode \epsfxsize=3.0in
  \epsffile{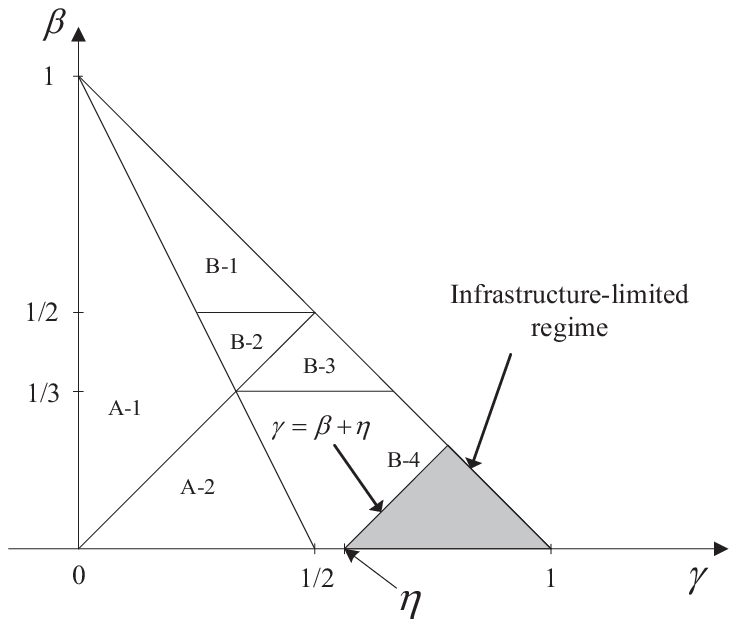}
  \caption{The infrastructure-limited regime with respect to $\beta$ and $\gamma$, where $\frac{1}{2}\leq\eta <1$.}
  \label{Fig:InfraLimitedRegimes-3}
\end{figure}
\fi

\begin{remark}
It is worth noting that the backhaul link rate required to
guarantee the maximum throughput scaling in
Theorem~\ref{Lem:TotalThroughputInfinite} regardless of system
parameters $m$ and $l$ is given by $\Omega(\sqrt{n})$ in Regimes
A, B-1, B-2, and B-3; and is given by $\Omega(n)$ in Regime B-4,
respectively. Hence, it turns out that the minimum rate of each
BS-to-BS link, $C_{\textrm{BS}}$, needed regardless of operating
regimes (or equivalently, the values of $m$ and $l$) is bounded by
$\Omega(n)$.
\end{remark}

\begin{remark}
Using the illustration of the infrastructure-limited regimes, it
would be interesting to examine whether or not the backhaul link
rate $R_{\textrm{BS}}$ needs to be scaled up when an operating
point varies according to scaling parameters $\beta$ and $\gamma$.
For example, let us assume that the network is at an operating
point $(\beta,\gamma)=(1/2,1/4)$ with $\eta=1/4$. In this case,
$R_{\textrm{BS}}$ does not need to be increased until $\gamma$
goes up to $1/2$.
\end{remark}

\section{Generalized Throughput Scaling With An Arbitrary Backhaul Link Rate}\label{SEC:GeneralizedScaling}

In this section, we shall derive the throughput scaling law for
the case where the rate of each BS-to-BS link scales at an
arbitrary rate relative to $n$, which generalizes the throughput
scaling in Theorem~\ref{Lem:TotalThroughputInfinite}.
In the following two lemmas, the throughput scaling results for
the ISH and IMH protocols operating under infinite-capacity
backhaul links shown in Lemmas~\ref{Lem:RateISH} and
\ref{Lem:RateIMH} are extended to a more general case having an
arbitrary backhaul link capacity, $R_{\textrm{BS}}$.

\begin{lemma}\label{Lem:AchievableRate-ISH}
Suppose that the ISH protocol is used in the hybrid network, where
the rate of each BS-to-BS link is limited by $R_{\textrm{BS}}$.
Then, the aggregate rate scaling is given by
\ifx \doubleColumn \undefined 
\begin{align} \label{Eq:ThroughputISH-BSlimited} 
    T_{n,\textrm{ISH}}
    =\Omega\left(\min\left\{ml\left(\frac{m}{n}\right)^{\alpha/2-1},m^2 R_{\textrm{BS}}, \frac{n}{\log
    n}R_{\textrm{BS}}\right\}\right),
\end{align}
\else
\begin{align} \label{Eq:ThroughputISH-BSlimited} 
    &T_{n,\textrm{ISH}}
    \nonumber\\
    &=\Omega\left(\min\left\{ml\left(\frac{m}{n}\right)^{\alpha/2-1},m^2 R_{\textrm{BS}}, \frac{n}{\log
    n}R_{\textrm{BS}}\right\}\right),
\end{align}
\fi where $m=n^\beta$.
\end{lemma}

\begin{IEEEproof}
From~(\ref{Eq:CBS-ISH-Rate-Tn}) in Lemma \ref{Lem:CBS-ISH}, for a
given transmission rate achieved by the ISH, $T_{n,\textrm{ISH}}$,
the minimum required rate of each BS-to-BS link,
$C_{\textrm{BS,ISH}}$, is given by
\begin{align*}
    C_{\textrm{BS,ISH}}= \left\{ \begin{array}{ll}
    \Omega \left( \frac{T_{n,\textrm{ISH}}\log n}{n}  \right) &{\rm{if~}}\frac{1}{2}\leq \beta < 1 \\
    \Omega \left( \frac{T_{n,\textrm{ISH}}}{m^2} \right) &{\rm{if~}}0\leq\beta<\frac{1}{2}. \\
    \end{array} \right.
\end{align*}
Hence, substituting a given backhaul link rate $R_{\textrm{BS}}$
into $C_{\textrm{BS,ISH}}$, the transmission rate
$T_{n,\textrm{ISH}}$ is given by
\ifx \doubleColumn \undefined 
\begin{align}\label{Eq:ThroughputISH-BSlimited-TwoConditions} 
    T_{n,\textrm{ISH}}
    = \left\{ \begin{array}{ll}
    \Omega \left( \min\left\{ml\left(\frac{m}{n}\right)^{\alpha/2-1},\frac{n}{\log n}R_{\textrm{BS}}\right\}  \right)
     &{\textrm{if~}}\frac{1}{2}\leq \beta < 1\\
    \Omega \left( \min\left\{ml\left(\frac{m}{n}\right)^{\alpha/2-1},m^2 R_{\textrm{BS}}\right\} \right)
    &{\textrm{if~}}0\leq\beta<\frac{1}{2}.
    \end{array}  \right.
\end{align}
\else
\begin{align}\label{Eq:ThroughputISH-BSlimited-TwoConditions} 
    T_{n,\textrm{ISH}}
    = \left\{ \begin{array}{ll}
    \Omega \left( \min\left\{ml\left(\frac{m}{n}\right)^{\alpha/2-1},\frac{n}{\log n}R_{\textrm{BS}}\right\}  \right)
    \\{\textrm{if~}}\frac{1}{2}\leq \beta < 1\\
    \Omega \left( \min\left\{ml\left(\frac{m}{n}\right)^{\alpha/2-1},m^2 R_{\textrm{BS}}\right\} \right)
    \\{\textrm{if~}}0\leq\beta<\frac{1}{2}.
    \end{array}  \right.
\end{align}
\fi From the fact that $\min\left\{\frac{n}{\log
n}R_{\textrm{BS}}, m^2 R_{\textrm{BS}}\right\}$ is given by $m^2
R_{\textrm{BS}}$ and $\frac{n}{\log n}R_{\textrm{BS}}$ as
$0\leq\beta<\frac{1}{2}$ and $\frac{1}{2}\leq\beta<1$,
respectively, (\ref{Eq:ThroughputISH-BSlimited-TwoConditions}) can
be simplified to (\ref{Eq:ThroughputISH-BSlimited}). Therefore,
$T_{n,\textrm{ISH}}$ is finally given by
(\ref{Eq:ThroughputISH-BSlimited}), which completes the proof of
Lemma~\ref{Lem:AchievableRate-ISH}.
\end{IEEEproof}

\ifx \doubleColumn \undefined 
\begin{figure}[t!] 
    \centering
  \leavevmode \epsfxsize=0.86\textwidth
  \epsffile{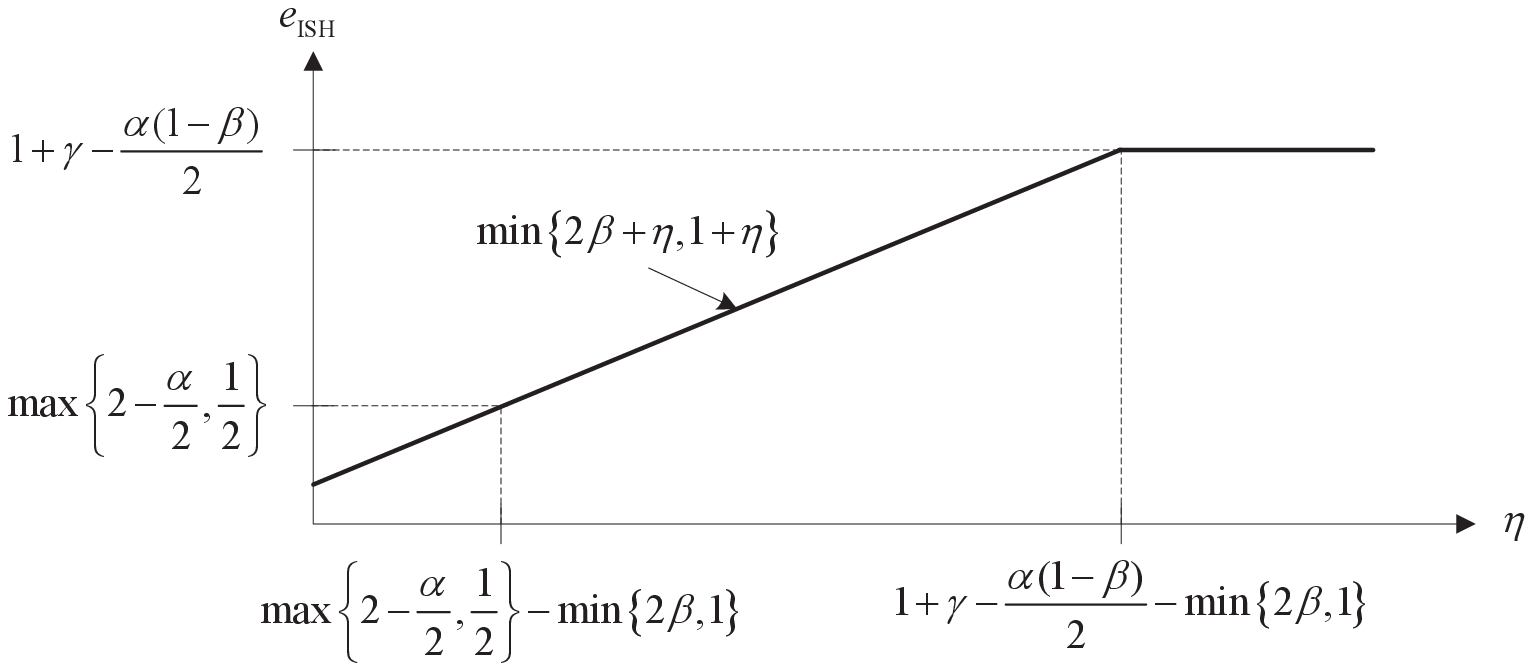}
  \caption{The throughput scaling exponent of the ISH protocol in the network with GreenInfra.}
  \label{Fig:ThroughputScalingISH}
\end{figure}
\else
\begin{figure}[t!] 
    \centering
  \leavevmode \epsfxsize=0.5\textwidth
  \epsffile{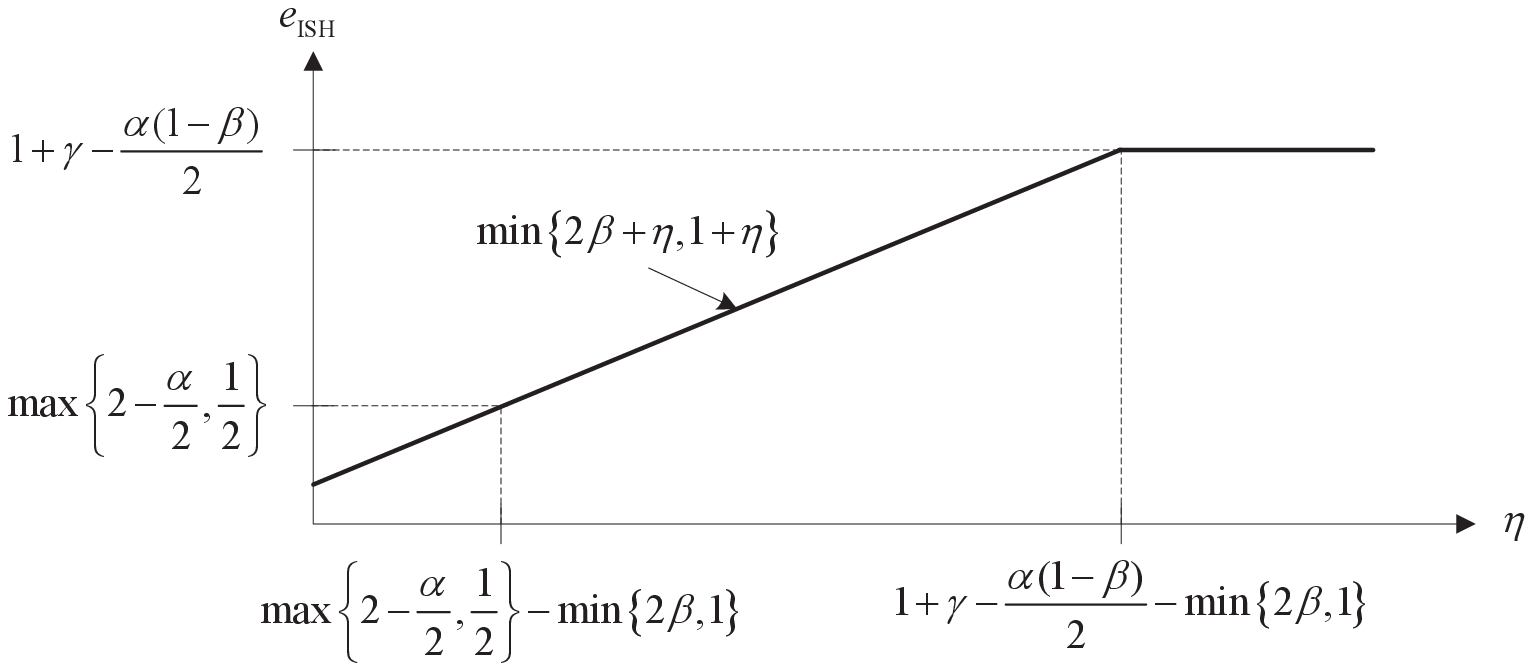}
  \caption{The throughput scaling exponent of the ISH protocol in the network with GreenInfra.}
  \label{Fig:ThroughputScalingISH}
\end{figure}
\fi

The first term of the right-hand side in
(\ref{Eq:ThroughputISH-BSlimited}) corresponds to the rate
achieved by the ISH protocol
in~(\ref{Eq:ThroughputISH-BSunlimited}) when the rate of the
BS-to-BS link is unlimited. The remaining two terms are the
throughput scalings supportable by BSs for an arbitrary
$R_{\textrm{BS}}$. The throughput scaling exponent of the ISH
protocol is shown in Fig.~\ref{Fig:ThroughputScalingISH}.

\begin{lemma}\label{Lem:AchievableRate-IMH}
Suppose that the IMH protocol is used in the hybrid network, where
the rate of each BS-to-BS link is limited by $R_{\textrm{BS}}$.
Then, the aggregate rate scaling is given by
\ifx \doubleColumn \undefined 
\begin{align}\label{Eq:ThroughputIMH-BSlimited} 
    T_{n,\textrm{IMH}}
    =
    \Omega \left( \min\left\{ ml,m\left(\frac{n}{m}\right)^{1/2-\epsilon},m^2R_{\textrm{BS}},
    \frac{ml}{\log n} R_{\textrm{BS}}
    ,\frac{m}{\log n}\left(\frac{n}{m}\right)^{1/2-\epsilon} R_{\textrm{BS}}\right\}
    \right),
\end{align}
\else
\begin{align}\label{Eq:ThroughputIMH-BSlimited} 
    T_{n,\textrm{IMH}}
    =
    \Omega \left( \min\left\{ ml,m\left(\frac{n}{m}\right)^{1/2-\epsilon},m^2R_{\textrm{BS}},
    \right.\right.
    \nonumber\\
    \left.\left.
    \frac{ml}{\log n} R_{\textrm{BS}}
    ,\frac{m}{\log n}\left(\frac{n}{m}\right)^{1/2-\epsilon} R_{\textrm{BS}}\right\}
    \right),
\end{align}
\fi where $\epsilon>0$ is an arbitrarily small constant.
\end{lemma}

\begin{IEEEproof}
From~(\ref{Eq:CBS-IMH-Rate-Tn}) in Lemma \ref{Lem:CBS-IMH}, for a
given transmission rate achieved by the IMH, $T_{n,\textrm{IMH}}$,
the minimum required rate of each BS-to-BS link,
$C_{\textrm{BS,IMH}}$, is given by
\ifx \doubleColumn \undefined 
\begin{align*} 
    C_{\textrm{BS,IMH}}
    = \left\{ \begin{array}{ll}
    \Omega\left(\frac{T_{n,\textrm{IMH}} }{ml}\log n\right)&
    \textrm{for Regime A-1}\\
    \Omega\left(\frac{T_{n,\textrm{IMH}} }{m^2}\right)&
    \textrm{for Regimes A-2 and B-4}\\
    \Omega\left(\frac{T_{n,\textrm{IMH}}}{m(n/m)^{1/2-\epsilon}}\log n\right)&
    \textrm{for Regimes B-1, B-2, and B-3}.
    \end{array} \right.
\end{align*}
\else
\begin{align*} 
    C_{\textrm{BS,IMH}}
    = \left\{ \begin{array}{ll}
    \Omega\left(\frac{T_{n,\textrm{IMH}} }{ml}\log n\right)&
    \textrm{for Regime A-1\\
    \Omega\left(\frac{T_{n,\textrm{IMH}} }{m^2}\right)&
    \textrm{for Regimes A-2 and B-4}\\
    \Omega\left(\frac{T_{n,\textrm{IMH}}}{m(n/m)^{1/2-\epsilon}}\log n\right)&
    \textrm{for Regimes B-1, B-2, and B-3}.
    \end{array} \right.
\end{align*}
\fi Hence, substituting a given backhaul link rate
$R_{\textrm{BS}}$ into $C_{\textrm{BS,IMH}}$, the transmission
rate $T_{n,\textrm{IMH}}$ is given by
\ifx \doubleColumn \undefined 
\begin{align} 
    T_{n,\textrm{IMH}}
    = \left\{ \begin{array}{ll}
    \Omega \left( \min\left\{ ml,m\left(\frac{n}{m}\right)^{1/2-\epsilon},\frac{ml}{\log n} R_{\textrm{BS}},
    \frac{m}{\log n}\left(\frac{n}{m}\right)^{1/2-\epsilon} R_{\textrm{BS}}\right\} \right)&\\
    {\textrm{for Regimes A-1, B-1, B-2, and B-3}} \\
    \Omega \left( \min\left\{ ml,m\left(\frac{n}{m}\right)^{1/2-\epsilon}, m^2R_{\textrm{BS}}\right\}  \right) \\
    {\textrm{for Regimes A-2 and B-4.}} \\
    \end{array} \right. \nonumber
\end{align}
\else
\begin{align} 
    &T_{n,\textrm{IMH}}
    \nonumber\\
    &= \left\{ \begin{array}{ll}
    \Omega \left( \min\left\{ ml,m\left(\frac{n}{m}\right)^{1/2-\epsilon},\frac{ml}{\log n} R_{\textrm{BS}},
    \right.\right.
    \nonumber\\
    \left.\left.
    ~~~~~~~~~~~~~\frac{m}{\log n}\left(\frac{n}{m}\right)^{1/2-\epsilon} R_{\textrm{BS}}\right\} \right)&\\
    {\textrm{for Regimes A-1, B-1, B-2, and B-3}} \\
    \Omega \left( \min\left\{ ml,m\left(\frac{n}{m}\right)^{1/2-\epsilon}, m^2R_{\textrm{BS}}\right\}  \right) \\
    {\textrm{for Regimes A-2 and B-4.}} \\
    \end{array} \right. \nonumber
\end{align}
\fi Since $\min\left\{\frac{ml}{\log n} R_{\textrm{BS}},
\frac{m}{\log n}\left(\frac{n}{m}\right)^{1/2-\epsilon}
R_{\textrm{BS}}\right\}$ is given by $\frac{ml}{\log n}
R_{\textrm{BS}}$ in Regime A-1 and $\frac{m}{\log
n}\left(\frac{n}{m}\right)^{1/2-\epsilon} R_{\textrm{BS}}$ in
Regimes B-1, B-2, and B-3, the transmission rate for Regimes A-1,
B-1, B-2, and B-3 can be expressed as a single term. Similarly,
$T_{n,\textrm{IMH}}$ is finally given by
(\ref{Eq:ThroughputIMH-BSlimited}), which completes the proof of
Lemma~\ref{Lem:AchievableRate-IMH}.
\end{IEEEproof}

\ifx \doubleColumn \undefined 
\begin{figure}[t!] 
    \centering
  \leavevmode \epsfxsize=0.86\textwidth
  \epsffile{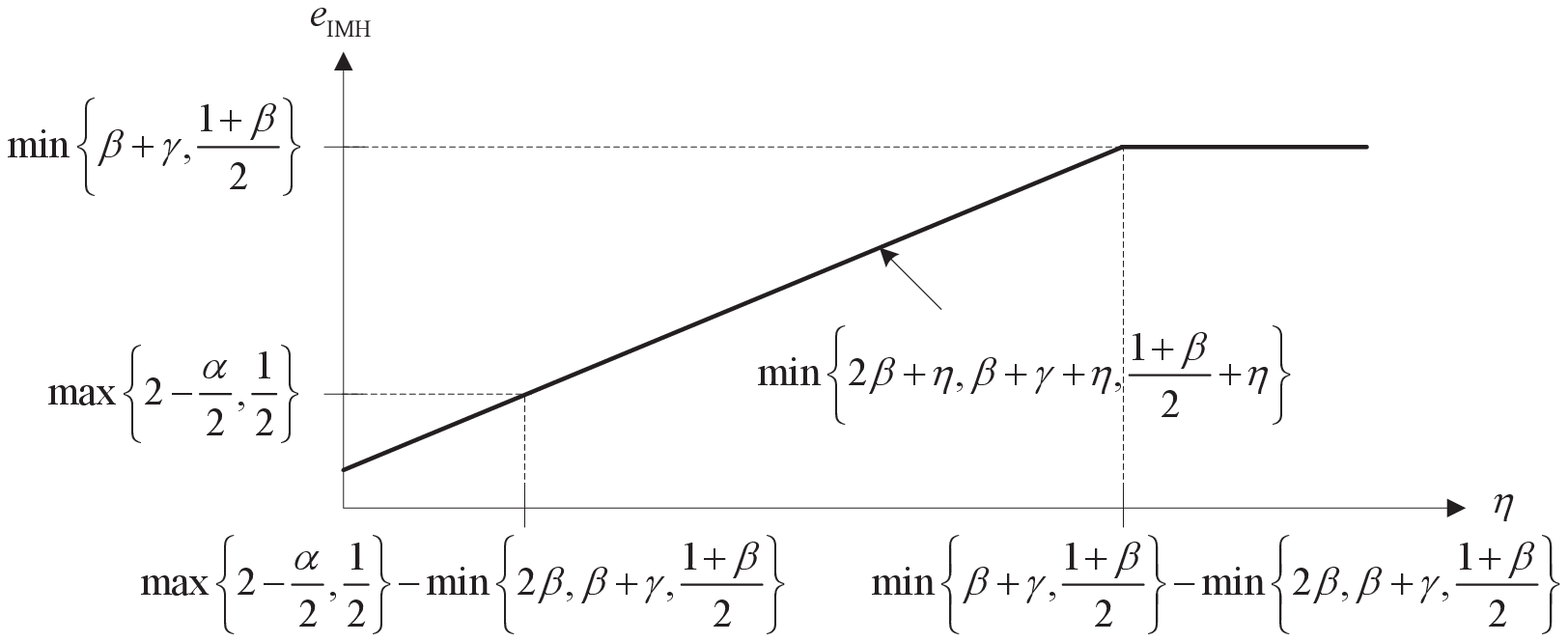}
  \caption{The throughput scaling exponent of the IMH protocol in the network with GreenInfra.}
  \label{Fig:ThroughputScalingIMH}
\end{figure}
\else
\begin{figure}[t!] 
    \centering
  \leavevmode \epsfxsize=0.5\textwidth
  \epsffile{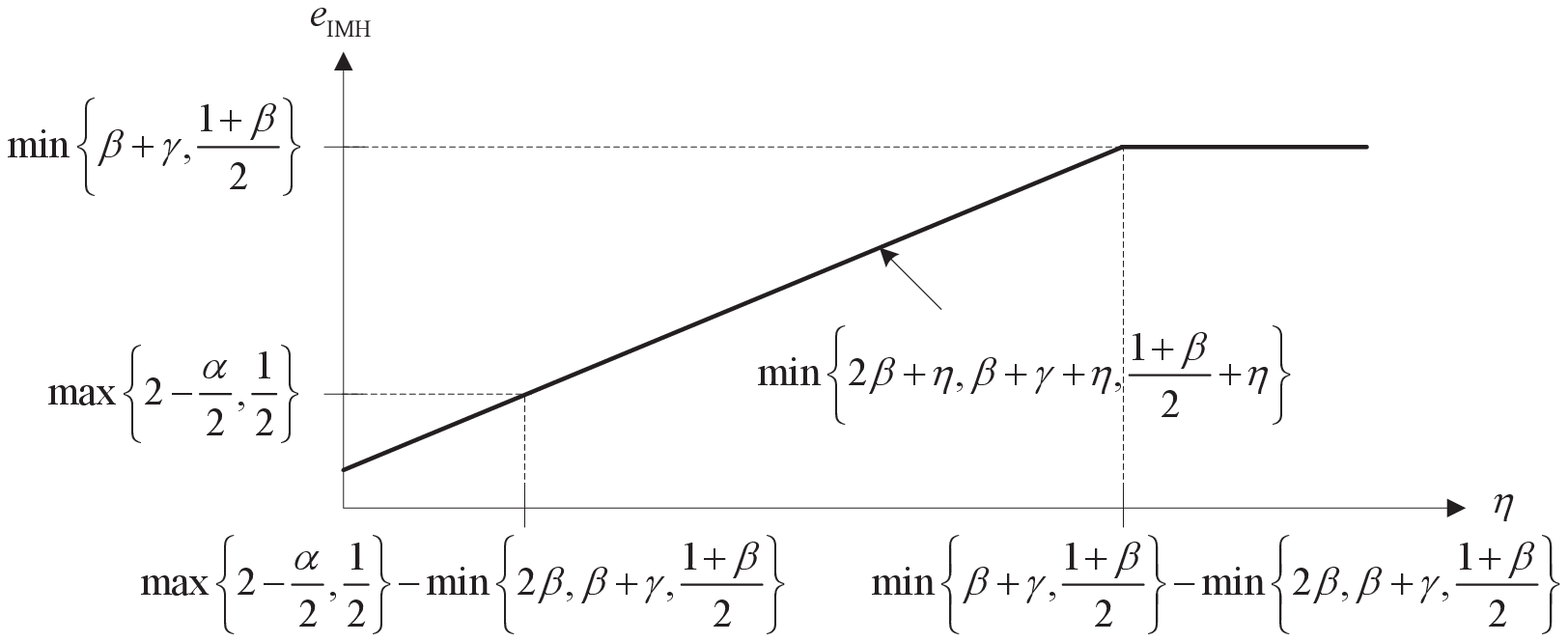}
  \caption{The throughput scaling exponent of the IMH protocol in the network with GreenInfra.}
  \label{Fig:ThroughputScalingIMH}
\end{figure}
\fi

The first two terms of the right-hand side in
(\ref{Eq:ThroughputIMH-BSlimited}) corresponds to the rate
achieved by the IMH protocol in
(\ref{Eq:ThroughputIMH-BSunlimited}) when the rate of backhaul
links is not limited. The remaining terms are the throughput that
can be supported via backhaul links. The throughput scaling
exponent of the IMH protocol is shown in
Fig.~\ref{Fig:ThroughputScalingIMH}.


Using Lemmas~\ref{Lem:AchievableRate-ISH}
and~\ref{Lem:AchievableRate-IMH}, we finally establish the
following theorem, which shows the aggregate throughput along with
the rate $R_{\textrm{BS}}$ of each BS-to-BS link.

\begin{theorem}\label{Thm:ThroughputRateLimitedBackhaul}
In the network with the backhaul link rate $R_{\textrm{BS}}$, the
aggregate throughput scales as
\begin{align}\label{Eq:ThroughputTotal-BSlimited}
T_n =\max\{T_{n,\textrm{ISH}},T_{n,\textrm{IMH}}\}.
\end{align}
\end{theorem}


It is easy to show that the transmission rate in
(\ref{Eq:ThroughputTotal-BSlimited}) is simplified to that in
(\ref{Eq:TotalThroughputInfinite}) as $R_{\textrm{BS}}\ge
C_{\textrm{BS}}$. Furthermore, we remark that the operating
regimes on the throughput scaling in
(\ref{Eq:ThroughputTotal-BSlimited}) with respect to $\beta$ and
$\gamma$ depend heavily on the value of $R_{\textrm{BS}}$.



\section{Numerical Evaluation}\label{SEC:NumericalResults}

\begin{table}[t]
    \centering
    \caption{Simulation environments}
    \label{Tab:Environments}
\begin{tabular}{|c|c|}
  \hline
  Parameter & Value\\
  \noalign{\hrule height 1.2pt}
  \hline
  \multirow{1}{*}{Average distance between nearest-neighbor nodes} & 100 m\\
  \hline
  \multirow{1}{*}{Transmit power} & -10 dBm\\
  \hline
  \multirow{1}{*}{Noise spectral density} & -174 dBm/Hz\\
  \hline
  \multirow{1}{*}{Noise figure} & 5 dB\\
  \hline
  \multirow{1}{*}{Bandwidth} & 40 MHz\\
  \hline
\end{tabular}
\end{table}

In this section, to validate our achievability result in
Section~\ref{SEC:GeneralizedScaling} in a realistic hybrid network
with GreenInfra, we perform computer simulations according to
finite values of the system parameters $n$, $m$, $l$, and
$R_{\textrm{BS}}$. From (\ref{Eq:ThroughputTotal-BSlimited}), the
aggregate throughput $T_n$ is numerically computed by taking the
maximum of $T_{n,\textrm{ISH}}$ and $T_{n,\textrm{IMH}}$ in a
certain operating regime. More specifically, the throughput
$T_{n,\textrm{ISH}}$ is numerically computed along with the ISH
protocol in Section~\ref{SEC:ISH}, for which the MMSE--SIC
postcoder and DPC-based precoder, shown in (\ref{EQ:MMSE}) and
(\ref{EQ:DPC}), respectively, are assumed. The throughput
$T_{n,\textrm{IMH}}$ is numerically computed along with the IMH
protocol in Section~\ref{SEC:IMH}. Simulation parameters are
summarized in TABLE~\ref{Tab:Environments}. A sufficient number of
nodes ($n\geq 256$) are deployed so that a large-scale network is
suitably modelled in practice. It is assumed that the average
distance between two nearest-neighbor nodes is 100 m (this
inter-node distance is sufficient to model an extended network
configuration). In our Monte-Carlo simulations, a sufficiently
large number of iterations are performed to average out the
aggregate throughput for both different node distributions and
channels (i.e., phases) under the given system parameters.

\begin{figure}[t!] 
    \centering
  \leavevmode \epsfxsize=0.68\textwidth
  \epsffile{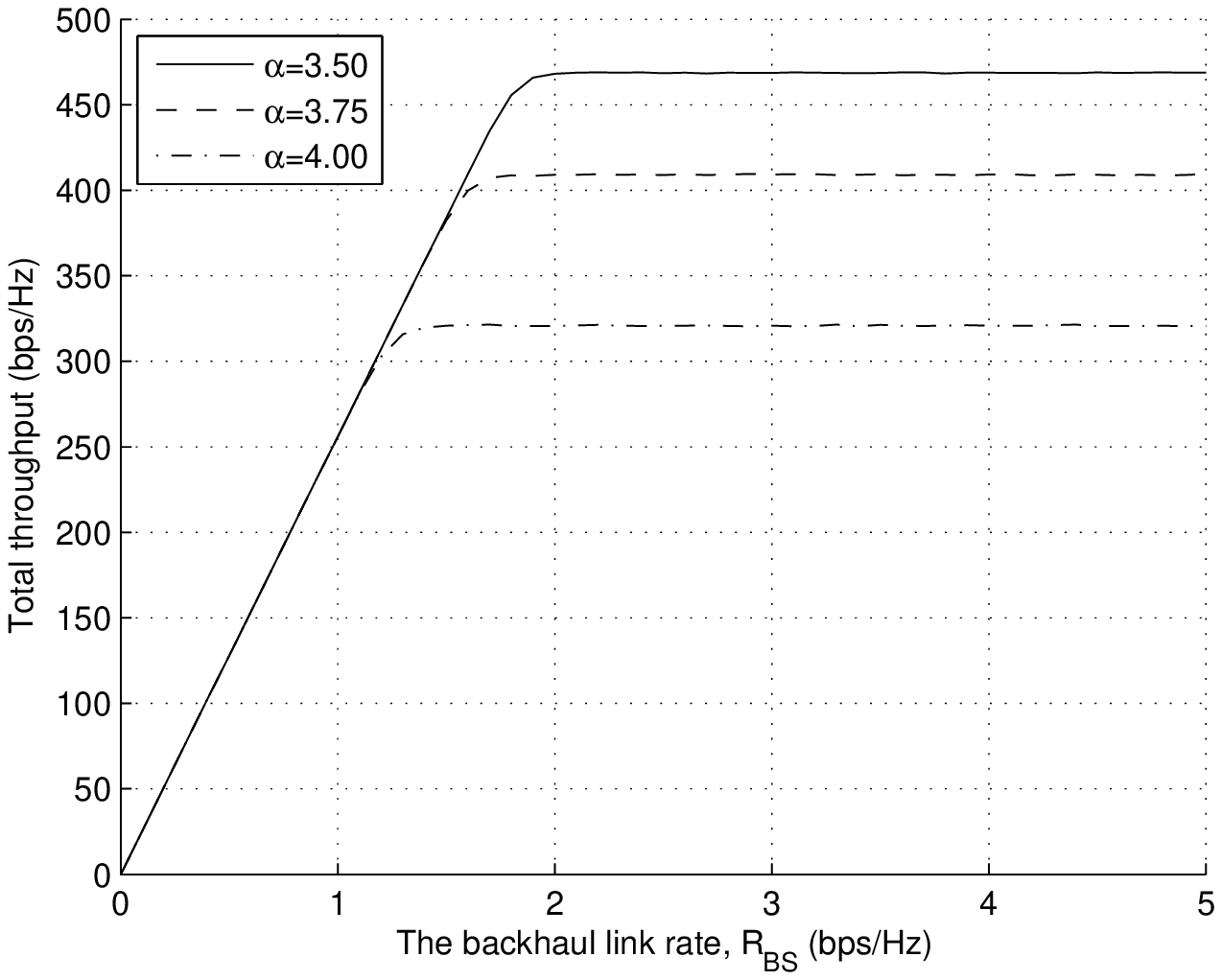}
  \caption{The aggregate throughput $T_n$ according to the backhaul link rate, $R_{\textrm{BS}}$, where $n$=1296, $m$=16, and
  $l$=4.}
  \label{Fig:Throughput-Rbs}
\end{figure}

The aggregate throughput $T_n$ versus the backhaul link rate
$R_{\textrm{BS}}$ is first evaluated in
Fig.~\ref{Fig:Throughput-Rbs}, where $n$=1296, $m$=16, $l$=4, and
$\alpha\in\{3.5, 3.75, 4.0\}$. From the figure, the following
interesting observations are made under our simulation
environments:

\begin{itemize}
\item The throughput $T_n$ is monotonically increasing with
$R_{\textrm{BS}}$ as $R_{\textrm{BS}}<C_{\textrm{BS}}$, where
$C_{\textrm{BS}}$ is given by $\{2.0, 1.8, 1.5\}$ for
$\alpha\in\{3.5,3.75,4.0\}$, respectively (see
TABLE~\ref{Table:comparison_alpha}).

\item For $R_{\textrm{BS}}<C_{\textrm{BS}}$ (i.e., the
infrastructure-limited regime), the throughput increases linearly
with $R_{\textrm{BS}}$, which is consistent with our analytical
prediction (refer to
(\ref{Eq:ThroughputISH-BSlimited})--(\ref{Eq:ThroughputTotal-BSlimited})).

\item For $R_{\textrm{BS}}\ge C_{\textrm{BS}}$, the throughput
performance is not improved with increasing $R_{\textrm{BS}}$.

\item The throughput $T_n$ gets significantly reduced as $\alpha$
increases (our scaling results indicate that the transmission rate
achieved by the ISH protocol is a non-increasing function of
$\alpha$).
\end{itemize}


\begin{table}[t]
    \centering
    \caption{Evaluation comparison according to the path-loss exponent $\alpha$, where $C_{\textrm{BS}}$ and the maximum $T_n$ are shown.}
\begin{tabular}{|c|c|c|}
  \hline
  $\alpha$ & $C_{\textrm{BS}}$ & Maximum value of $T_n$ \\
  \noalign{\hrule height 1.2pt}
  \hline
  3.5 & 2.0 & 468.7\\
  \hline
  3.75 & 1.8 & 409.3\\
  \hline
  4.0 & 1.5 & 322.6\\
  \hline
\end{tabular} \label{Table:comparison_alpha}
\end{table}

\begin{figure}[t!] 
    \centering
  \leavevmode \epsfxsize=0.68\textwidth
  \epsffile{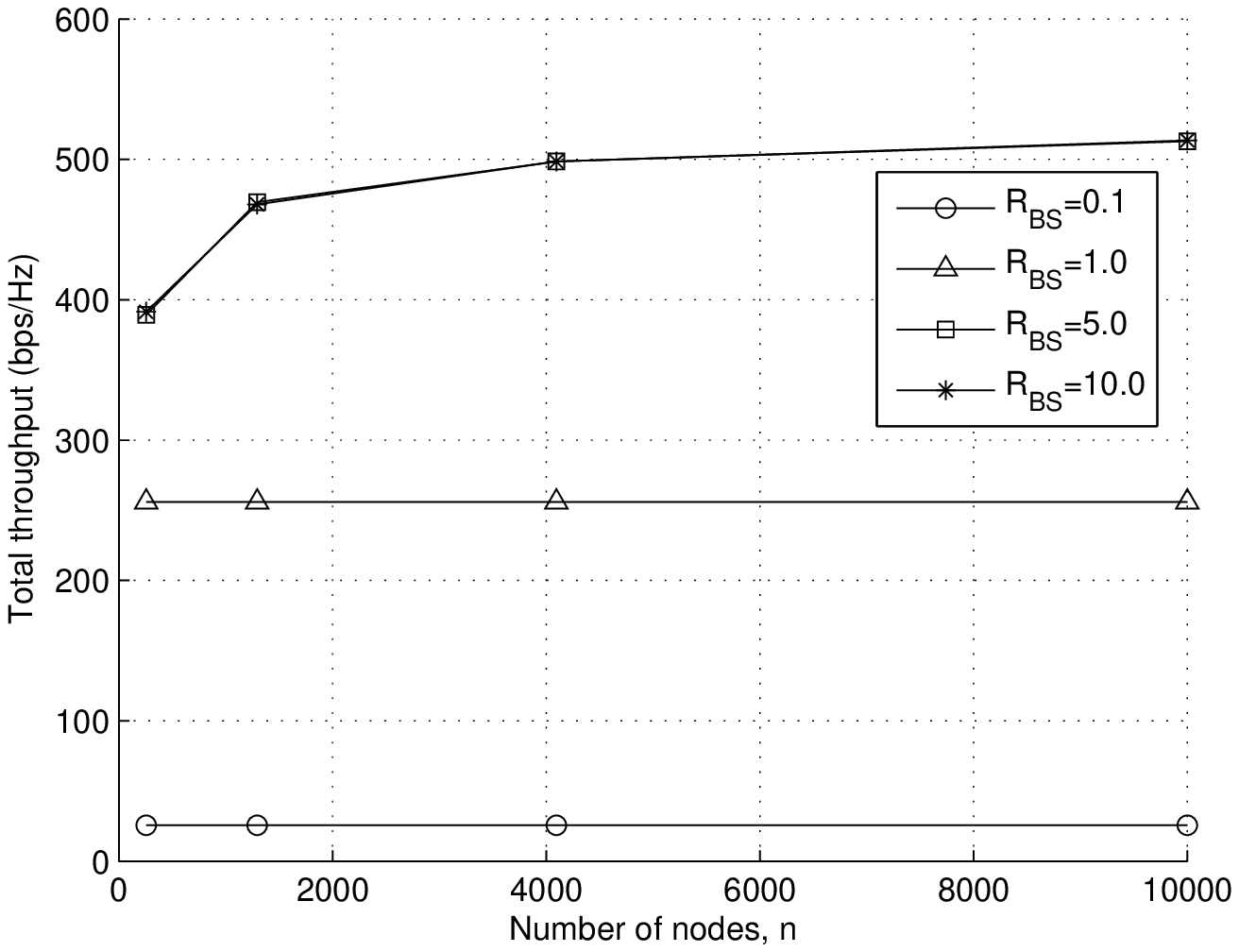}
  \caption{The aggregate throughput $T_n$ according to the number of nodes, $n$, where $\alpha$=3.5, $m$=16, and
  $l$=4.}
  \label{Fig:Throughput-Nodes}
\end{figure}

We remark that the most cost-effective way in designing the
GreenInfra is to set $R_{\textrm{BS}}=C_{\textrm{BS}}$ while
maintaining the maximum throughput we can hope for (i.e., the same
throughput as in the infinite-capacity backhaul link case). In
addition, the throughput $T_n$ versus the number of nodes,
$n\in\{256, 1296, 4096, 10000\}$, is evaluated in
Fig.~\ref{Fig:Throughput-Nodes}, where $\alpha$=3.5, $m$=16,
$l$=4, and $R_{\textrm{BS}}\in\{0.1, 1.0, 5.0, 10.0\}$. From the
figure, the following observations are found:

\begin{itemize}
\item It is numerically found that the ISH has a better
performance than that of the IMH under the aforementioned
simulation environment. \item For this reason, when
$R_{\textrm{BS}}<C_{\textrm{BS}}$ (e.g., $R_{\textrm{BS}}=0.1,
1.0$ in the figure), the throughput $T_n$ is not improved with
increasing $n$ since the rate of the ISH is limited to $m^2
R_{\textrm{BS}}$ (but not to $n R_{\textrm{BS}}$) from
(\ref{Eq:ThroughputISH-BSlimited}). \item For $R_{\textrm{BS}}\ge
C_{\textrm{BS}}$, the same throughput as in the infinite-capacity
backhaul link case is achieved. Thus, the throughput for
$R_{\textrm{BS}}=5.0$ is identical to that for
$R_{\textrm{BS}}=10.0$, which approaches the maximum we can hope
for.
\end{itemize}

\begin{figure}[t!] 
    \centering
  \leavevmode \epsfxsize=0.68\textwidth
  \epsffile{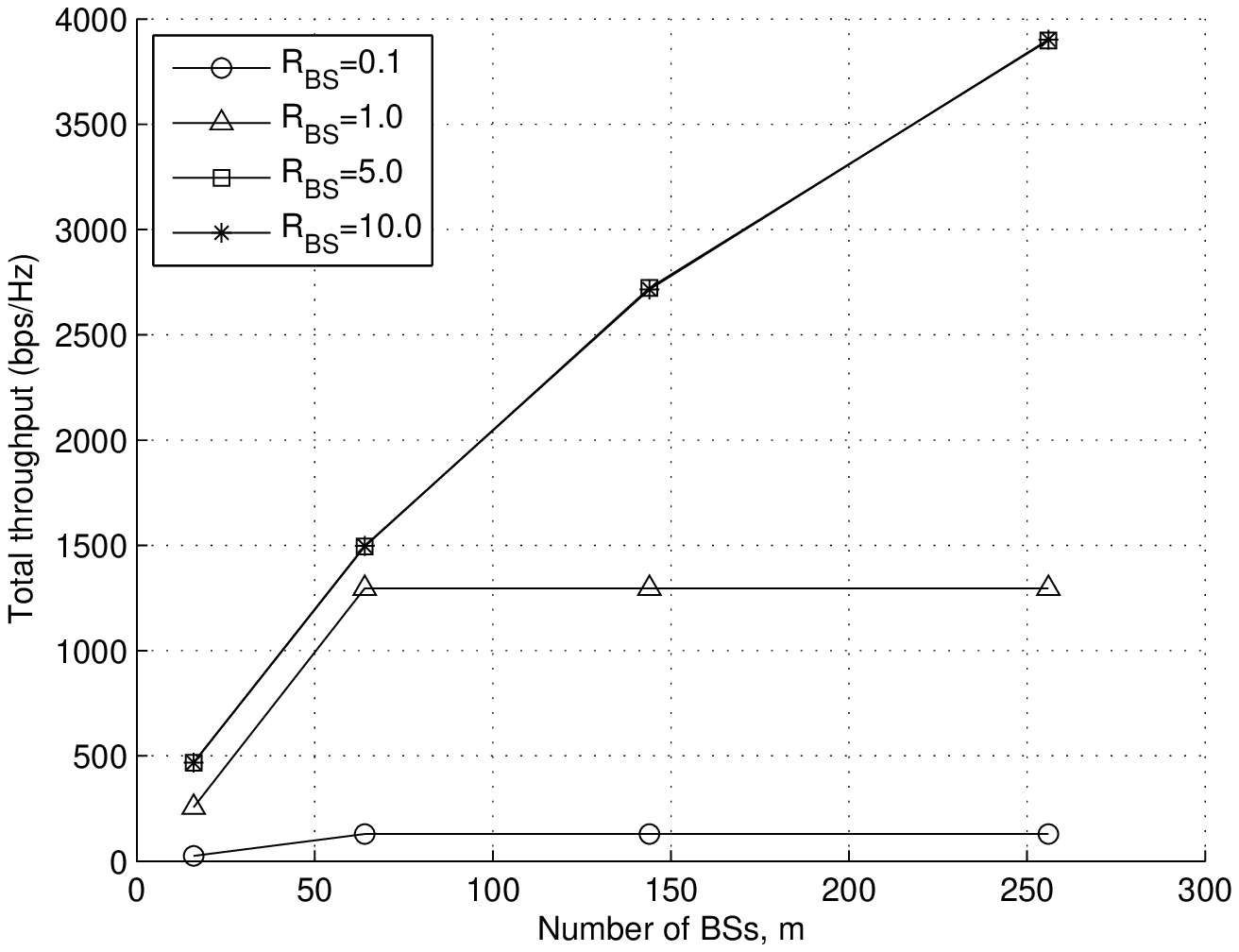}
  \caption{The aggregate throughput $T_n$ according to the number of BSs, $m$, where $\alpha$=3.5, $n$=1296, and
  $l$=4.}
  \label{Fig:Throughput-BSs}
\end{figure}

The throughput $T_n$ versus the number of BSs, $m\in\{16, 64, 144,
256\}$, is also evaluated in Fig.~\ref{Fig:Throughput-BSs}, where
$\alpha$=3.5, $n$=1296, $l$=4, and $R_{\textrm{BS}}\in\{0.1, 1.0,
5.0, 10.0\}$. From the figure, the following observations are
found:

\begin{itemize}
\item It is numerically found that the ISH protocol is dominant
under the simulation environment. \item We first focus on the
infrastructure-limited regime (i.e.,
$R_{\textrm{BS}}<C_{\textrm{BS}}$). In this regime, for small $m$
(e.g., $m=16$), the throughput $T_n$ is limited to
$m^2R_{\textrm{BS}}$ from (\ref{Eq:ThroughputISH-BSlimited}). When
$m$ increases, the throughput is now limited to
$nR_{\textrm{BS}}$, thereby resulting in no performance
improvement for $m\in\{64, 144, 256\}$. \item Let us turn to the
case where $R_{\textrm{BS}}\ge C_{\textrm{BS}}$. The two backhaul
link rates $R_{\textrm{BS}}=5.0$ and $R_{\textrm{BS}}=10.0$ lead
to the same throughput. It is also seen that the throughput
monotonically increases with $m$ due to the fact that its scaling
is given by $ml\left(\frac{m}{n}\right)^{\alpha/2-1}$ from
(\ref{Eq:ThroughputISH-BSlimited}).
\end{itemize}

\begin{figure}[t!] 
    \centering
  \leavevmode \epsfxsize=0.68\textwidth
  \epsffile{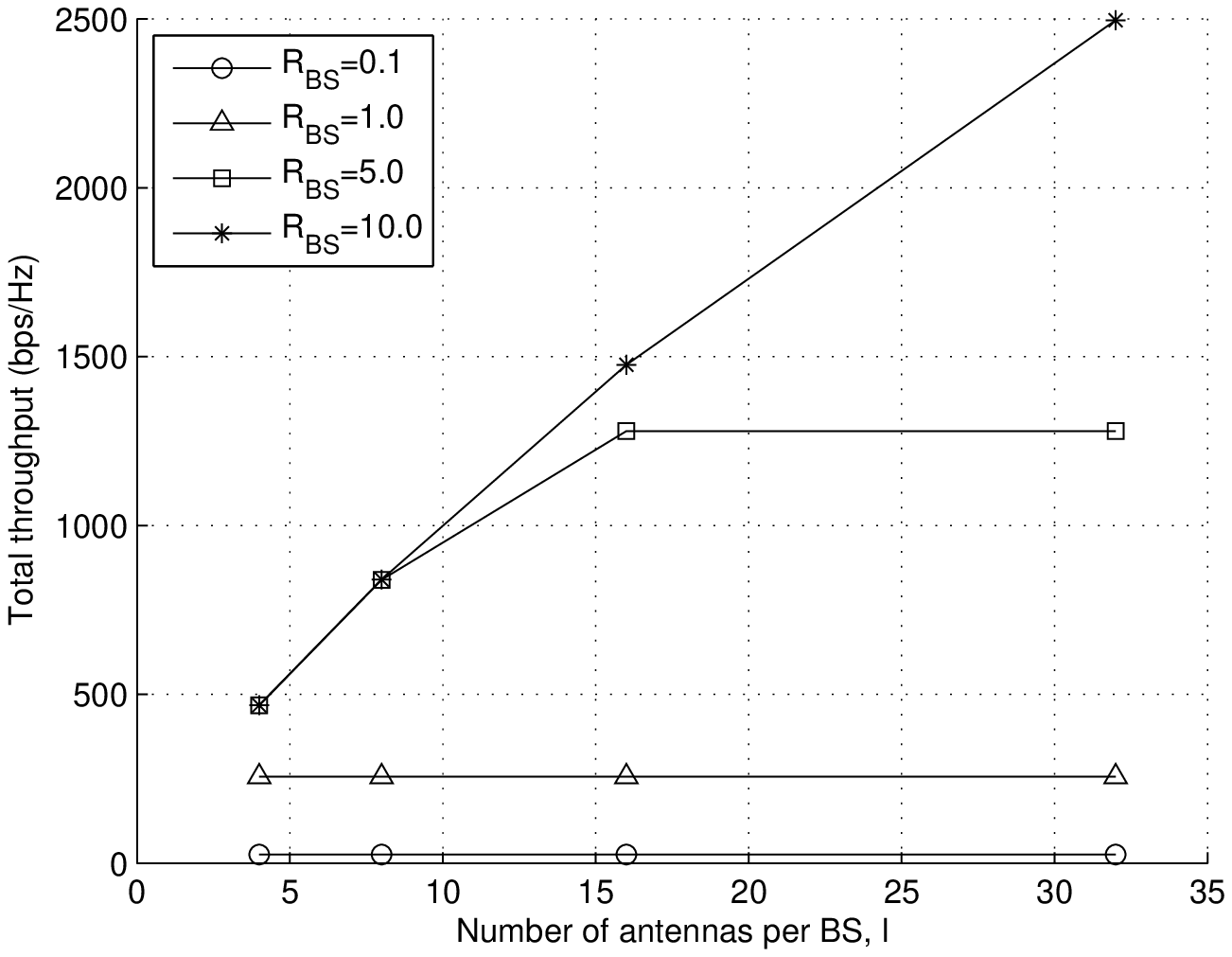}
  \caption{The aggregate throughput $T_n$ according to the number of antennas per BS, $l$, where $\alpha$=3.5, $n$=1296, and
  $m$=16.}
  \label{Fig:Throughput-Antennas}
\end{figure}

Finally, the throughput $T_n$ versus the number of antennas per
BS, $l\in\{4, 8, 16, 32\}$, is evaluated in
Fig.~\ref{Fig:Throughput-Antennas}, where $\alpha$=3.5, $n$=1296,
$m$=16, and $R_{\textrm{BS}}\in\{0.1, 1.0, 5.0, 10.0\}$. From the
figure, the following observations are found:

\begin{itemize}
\item Similarly, the ISH protocol is dominant under the simulation
environment.
 \item As mentioned before, when
$R_{\textrm{BS}}$ is small (e.g., $R_{\textrm{BS}}=0.1, 1.0$), the
throughput is limited to $m^2 R_{\textrm{BS}}$ and thus does not
depend on $l$. \item We recall Theorem~\ref{Thm:RequiredRate},
which implies that $C_{\textrm{BS}}$ is given by
$n^{\gamma-\beta}$ in Regimes A-2, B-3, and B-4. From
Fig.~\ref{Fig:OperatingRegimesFiniteBScapacity}, one can see that
as $l$ increases, our network approaches one of the three regimes
above. \item We consider the case where $R_{\textrm{BS}}=5.0$. In
this case, the throughput $T_n$ increases with $l$ for small $l$,
which means that $R_{\textrm{BS}}\ge C_{\textrm{BS}}$. However,
$T_n$ is no longer improved beyond a certain value of $l$ ($l=16$
in the figure), which in turn implies that our network is
infrastructure-limited.
\end{itemize}

\section{Concluding Remarks}\label{SEC:Conclusion}
The large-scale hybrid network with an arbitrary rate scaling of
each backhaul link based on the two infrastructure-supported
routing protocols, ISH and IMH, was completely characterized. For
the two-dimensional operating regimes with respect to the number
of BSs, $m$, and the number of antennas at each BS, $l$, the
minimum required rate of each BS-to-BS link, $C_{\textrm{BS}}$,
was derived to achieve the optimal throughput scaling for a
cost-effective backhaul solution. The infrastructure-limited
regime was also explicitly identified with respect to the three
scaling parameters: $m$, $l$, and the backhaul link rate
$R_{\textrm{BS}}$. Furthermore, the generalized throughput scaling
was derived including the case for which the backhaul link rate
scales slower than the minimum required rate $C_{\textrm{BS}}$.
Finally, our scaling results were validated for finite values of
$n$, $m$, $l$ and $R_{\textrm{BS}}$ via computer simulations. The
numerical results were shown to be consistent with our analysis.

\appendix
\section{Appendix}
\renewcommand\theequation{\Alph{section}.\arabic{equation}}
\setcounter{equation}{0}

\subsection{Proof of Lemma~\ref{Lem:NumPairs}} \label{PF:NumPairs}
Since each node chooses its destination randomly and independently
in a cell among $n^b$ cells, the number of destinations in the
$k$th cell whose source nodes are in the $i$th cell, $X_{ki}$ does
not depend on $k$ and $i$. Hence, we derive $X_{ki}$ for arbitrary
$k$ and $i$ as in the following. The probability that the
destination node of the $j$th source node in the $i$th cell is in
the $k$th cell follows the Bernoulli distribution with
probabilities $\Pr\{B_j=1\}=1/n^b$ and $\Pr\{B_j=0\}=1-1/n^b$,
where $B_j$ is the Bernoulli random variable corresponding to the
$j$th source node. Note that $B_j$'s are independent over $j$
since each node chooses its destination independently. If $a\leq
b$, then the probability that the number of destination nodes in
the $k$th cell whose source nodes are in the $i$th cell is greater
than or equal to $\log n$ is given by
\ifx \doubleColumn \undefined 
\begin{align}\label{Eq:Xki-Bernoulli} 
   & \Pr\left\{\sum_{j=1}^{n^a}B_j \geq \log n\right\}
   =\Pr\left\{e^{s \sum_{j=1}^{n^a}B_j}\geq e^{s\log n}\right\}
   \nonumber\\
   &\leq\left(\mathbb{E}\left[e^{s \sum_{j=1}^{n^a}B_j}\right]\right)e^{-s\log n}
   =\left(\mathbb{E}\left[e^{sB_1}\right]\right)^{n^a}e^{-s\log n}
   =\left(e^s\frac{1}{n^b}+\left(1-\frac{1}{n^b}\right)\right)^{n^a}e^{-s\log n}
   \nonumber\\
   &\leq e^{-n^{a-b}(1-e^s)-s\log n},
\end{align}
\else
\begin{align} 
   & \Pr\left\{\sum_{j=1}^{n^a}B_j \geq \log n\right\} \nonumber\\
    &= \Pr\left\{e^{s \sum_{j=1}^{n^a}B_j}\geq e^{s\log n}\right\}
    \nonumber\\
    &\leq\left(\mathbb{E}[e^{sB_1}]\right)^{n^a}e^{-s\log n}
\nonumber\\&
=\left(e^s\frac{1}{n^b}+\left(1-\frac{1}{n^b}\right)\right)^{n^a}e^{-s\log
n}
    \nonumber\\
    &\leq e^{-n^{a-b}(1-e^s)-s\log n},
\end{align}
\fi where $s$ is a positive real value and $n^a$ is the number of
source nodes in a cell. Here, the first inequality follows from
the Markov inequality, i.e., $\Pr\{X\geq a\}\leq\mathbb{E}[X]/a$
where $X$ is any nonnegative random variable and $a>0$. The last
inequality in (\ref{Eq:Xki-Bernoulli}) follows from $1+x\leq e^x$
for all $x$. Since the right side of the last inequality above
tends to zero as $n$ goes to infinity, $X_{ki}$ is given by
$O(\log n)$ whp as $n$ goes to infinity if $a\leq b$.

Using \cite[Lemma 1]{ShinJeonDevroyeVuChungLeeTarokh:08}, it can
be shown that, if $a>b$, then $X_{ki}$ is between
$((1-\delta_0)n^{a-b},(1+\delta_0)n^{a-b})$, with probability
larger than
\begin{align}\label{Eq:Prob-NodeDist}
    1-n^be^{-\Delta(\delta_0)n^{a-b}},
\end{align}
where $\Delta(\delta_0)=(1+\delta_0)\ln(1+\delta_0)-\delta_0$ for
$0<\delta_0<1$ independent of $n$. The probability in
(\ref{Eq:Prob-NodeDist}) tends to one as $n$ increases. Hence,
$X_{ki}$ is given by $\Theta(n^{a-b})$ if $a>b$. Therefore, the
number of destinations in the $k$th cell whose source nodes are in
the $i$th cell is given by (\ref{Eq:NumSDpairs}), which completes
the proof of Lemma~\ref{Lem:NumPairs}.

\subsection{Proof of Lemma~\ref{Lem:CBS-ISH}} \label{PF:CBS-ISH}
Since there are $m=n^{\beta}$ cells in the network and
$n/m=n^{1-\beta}$ nodes in each cell whp, using Lemma
\ref{Lem:NumPairs}, we have
\begin{align}
    X_{ki}=\left\{\begin{array}{ll}
    O\left(\log n\right) &\textrm{if $\beta\geq\frac{1}{2}$}\\
    O\left(\frac{n}{m^2}\right) &\textrm{if $\beta<\frac{1}{2}$}.\\
    \end{array}
    \right. \nonumber
\end{align}
The operating regime in which $\beta\geq\frac{1}{2}$ is satisfied
corresponds to Regime B-1, while the operating regime in which
$\beta<\frac{1}{2}$ is satisfied corresponds to Regimes B-2, B-3,
and B-4. Hence, $X_{ki}$ is given by (\ref{Eq:BoundPacketsISH}).
Since each source transmits at a rate $T_{n,\textrm{ISH}}/n$ and
the number of S--D pairs between two BSs is bounded by
(\ref{Eq:BoundPacketsISH}), the required rate of each BS-to-BS
link for the ISH protocol is given by
\begin{align}\label{Eq:CBS-ISH-Rate-Tn}
    C_{\textrm{BS,ISH}}= \left\{ \begin{array}{ll}
    \Omega \left( \frac{T_{n,\textrm{ISH}}\log n}{n}  \right)&\textrm{for Regime B-1} \\
    \Omega \left( \frac{T_{n,\textrm{ISH}}}{m^2} \right) &\textrm{for Regimes B-2, B-3, and B-4}.
    \end{array} \right.
\end{align}
Since the transmission rate achieved by the ISH protocol with
infinite backhaul link rate is
$T_{n,\textrm{ISH}}=ml\left(\frac{m}{n}\right)^{\alpha/2-1}$ and
it is a decreasing function of $\alpha$, the minimum required rate
that supports any value of $\alpha$ can be obtained by
substituting $\alpha=2$ into (\ref{Eq:CBS-ISH-Rate-Tn}) as
follows:
\ifx \doubleColumn \undefined 
\begin{align*} 
    C_{\textrm{BS,ISH}}= \left\{ \begin{array}{ll}\
    \Omega \left(\frac{ml\log n}{n} \right)
    =\Omega \left( (\log n) n^{\beta+\gamma-1}\right)
    &\textrm{for Regime B-1} \\
    \Omega \left( \frac{l}{m}\right)
    =\Omega \left( n^{\gamma-\beta}   \right)
    &\textrm{for Regimes B-2, B-3, and B-4,}
    \end{array} \right.
\end{align*}
\else
\begin{align*} 
    C_{\textrm{BS,ISH}}= \left\{ \begin{array}{ll}\
    \Omega \left( \frac{ml\log n}{n} \right)
    =\Omega \left( (\log n) n^{1-\frac{1-\gamma}{\beta}}\right)
    \\\textrm{for Regime B-1} \\
    \Omega \left( \frac{l}{m}\right)
    =\Omega \left( n^{\gamma-\beta}\right)
    \\\textrm{for Regimes B-2, B-3, and B-4,}
    \end{array} \right.
\end{align*}
\fi which completes the proof of Lemma~\ref{Lem:CBS-ISH}.

\subsection{Proof of Lemma~\ref{Lem:CBS-IMH}} \label{PF:CBS-IMH}
The number of sources in the $i$th cell is denoted by
$\mathcal{X}_i$. The event that $\mathcal{X}_i$ is between
$((1-\delta_0)n/m,(1+\delta_0)n/m)$ for all $i\in\{1,\ldots,m\}$
is denoted by $A_{\mathcal{X}}$, where $0<\delta_0<1$ is a
constant independent of $n$. The probability that the destination
node of a transmitting source node is in the $k$th cell follows
the Bernoulli distribution with probabilities $\Pr\{B_j=1\}=1/m$
and $\Pr\{B_j=0\}=1-1/m$, where $B_j$ is a Bernoulli random
variable corresponding to the $j$th transmitting source node and
is independent over $j$. The probability that $X_{ki}$ is less
than $a$ is given by
\ifx \doubleColumn \undefined 
\begin{align}\label{Eq:ProbXki} 
    &\Pr\{X_{ki}<a\textrm{ for all }i, k\in\{1,\ldots,m\}\}
    \geq \Pr\{A_{\mathcal{X}}\}\Pr\{X_{ki}<a\textrm{ for all }i,k|A_{\mathcal{X}}\}
    \nonumber\\
    &\geq \Pr\{A_{\mathcal{X}}\}\left(1-m^2\Pr\left\{\sum_{j=1}^{\min\{l,(n/m)^{1/2-\epsilon}\}}B_j\geq
    a\right\}\right),
\end{align}
\else
\begin{align}\label{Eq:ProbXki} 
    &\Pr\{\mathcal{X}<a\textrm{ for all }i, k\in\{1,\ldots,m\}\}
    \nonumber\\
    &\geq \Pr\{A_{\mathcal{X}}\}\Pr\{\mathcal{X}<a\textrm{ for all }i,k|A_{\mathcal{X}}\}
    \nonumber\\
    &\geq \Pr\{A_{\mathcal{X}}\}\left(1-m^2\Pr\left\{\sum_{j=1}^{\min\{l,(n/m)^{1/2-\epsilon}\}}B_j\geq
    a\right\}\right),
\end{align}
\fi where $\Pr\{A_{\mathcal{X}}\}\geq 1-
n^{\beta}e^{-\Delta(\delta_0)n^{1-\beta}}$ and
$\Delta(\delta_0)=(1+\delta_0)\log(1+\delta_0)-\delta_0$. In
(\ref{Eq:ProbXki}), the second inequality holds since the union
bound is applied over all $i,k\in\{1,\ldots,m\}$. Note that
$\Pr\{A_{\mathcal{X}}\}$ converges to one as $n$ goes to infinity.

In the following, the scaling law of $X_{ki}$ is derived for four
cases with respect to $\beta$ and $\gamma$.

1) Regime A-1 $(\beta+2\gamma\leq 1 \textrm{~and~}
\gamma\leq\beta)$: Since $\beta+2\gamma\leq1$, the number of
simultaneously transmitting nodes in each cell is $l$. From Lemma
\ref{Lem:NumPairs}, we set $a=\log n$ because $\gamma\leq \beta$.
Then, we have
\ifx \doubleColumn \undefined 
\begin{align}\label{Eq:RegimeE-ProbB} 
    &\Pr\left\{\sum_{j=1}^{l}B_j\geq\log n\right\}
    =\Pr\left\{e^{s\sum_{j=1}^{l}B_j}\geq e^{s\log n}\right\}
    \leq (\mathbb{E}[e^{sB_1}])^{l}e^{-s\log n}
    \nonumber\\
    &=\left(e^s\frac{1}{m}+\left(1-\frac{1}{m}\right)\right)^{l}e^{-s\log n}
    \leq e^{-(1/m)(1-e^s)l}e^{-s\log n}
    =e^{-(1-e^s)l/m-s\log n},
\end{align}
\else
\begin{align}\label{Eq:RegimeE-ProbB} 
    &\Pr\left\{\sum_{j=1}^{l}B_j\geq\log n\right\} \nonumber\\
    &=\Pr\left\{e^{s\sum_{j=1}^{l}B_j}\geq e^{s\log n}\right\}
    \nonumber\\
    &\leq (\mathbb{E}[e^{sB_1}])^{l}e^{-s\log n}
    \nonumber\\
    &=\left(e^s\frac{1}{m}+\left(1-\frac{1}{m}\right)\right)^{l}e^{-s\log n}
\nonumber\\&    \leq e^{-(1/m)(1-e^s)l}e^{-s\log n}
    \nonumber\\
    &=e^{-(1-e^s)l/m-s\log n},
\end{align}
\fi where $s$ is a positive real value. Using
(\ref{Eq:RegimeE-ProbB}), the probability in (\ref{Eq:ProbXki}) is
given by
\ifx \doubleColumn \undefined 
\begin{align} 
    \Pr\{X_{ki}<\log n\textrm{ for all }i, k\in\{1,\ldots,m\}\}
    &\geq (1-n^\beta e^{-\Delta(\delta_0)n^{1-\beta}})
    (1-m^2 e^{-(1-e^s)l/m-s\log n})
    \nonumber\\
    &=(1-n^\beta e^{-\Delta(\delta_0)n^{1-\beta}})
    (1-e^{2\beta \log n -n^{\gamma-\beta}(1-e^s)-s\log n}), \nonumber
\end{align}
\else
\begin{align} 
    &\Pr\{\mathcal{X}<\log n\textrm{ for all }i, k\in\{1,\ldots,m\}\}
    \nonumber\\
    &\geq (1-n^\beta e^{-\Delta(\delta_0)n^{1-\beta}})
    (1-m^2 e^{-(1-e^s)l/m-s\log n})
    \nonumber\\
    &=(1-n^\beta e^{-\Delta(\delta_0)n^{1-\beta}})
    (1-e^{2\beta \log n -n^{\gamma-\beta}(1-e^s)-s\log n}), \nonumber
\end{align}
\fi which converges to one as $n$ goes to infinity by choosing
$s=(2+\delta_0)\beta$, where $m^2=e^{2\beta\log n}$.

2) Regime A-2 $(\beta+2\gamma\leq 1 \textrm{~and~} \gamma>\beta)$:
Since $\beta+2\gamma\leq1$, the number of simultaneously
transmitting nodes in each cell is $l$. From Lemma
\ref{Lem:NumPairs}, we set $a=(1+\delta_0)\frac{l}{m}$ because
$\gamma> \beta$. Then, we have
\ifx \doubleColumn \undefined 
\begin{align}\label{Eq:RegimeF-ProbB} 
    &\Pr\left\{\sum_{j=1}^{l}B_j\geq (1+\delta_0)\frac{l}{m}\right\}
    =\Pr\left\{e^{s\sum_{j=1}^{l}B_j}\geq e^{s(1+\delta_0)l/m}\right\}
    \leq (\mathbb{E}[e^{sB_1}])^{l}e^{-s(1+\delta_0)l/m}
    \nonumber\\
    &=\left(e^s\frac{1}{m}+(1-\frac{1}{m})\right)^{l}e^{-s(1+\delta_0)l/m}
    \leq e^{-(1/m)(1-e^s)l}e^{-s(1+\delta_0)l/m}
    =e^{-(1-e^s+s(1+\delta_0))l/m},
\end{align}
\else
\begin{align}\label{Eq:RegimeF-ProbB} 
    &\Pr\left\{\sum_{j=1}^{l}B_j\geq (1+\delta_0)\frac{l}{m}\right\} \nonumber\\
    &=\Pr\left\{e^{s\sum_{j=1}^{l}B_j}\geq e^{s(1+\delta_0)l/m}\right\}
    \nonumber\\
    &\leq (\mathbb{E}[e^{sB_1}])^{l}e^{-s(1+\delta_0)l/m}
    \nonumber\\&=\left(e^s\frac{1}{m}+(1-\frac{1}{m})\right)^{l}e^{-s(1+\delta_0)l/m}
    \nonumber\\
    &\leq e^{-(1/m)(1-e^s)l}e^{-s(1+\delta_0)l/m}
    \nonumber\\&=e^{-(1-e^s+s(1+\delta_0))l/m},
\end{align}
\fi where $s$ is a positive real value. Using
(\ref{Eq:RegimeF-ProbB}), the probability in (\ref{Eq:ProbXki}) is
given by
\begin{align}
    &\Pr\{X_{ki}<(1+\delta_0)l/m\}\textrm{ for all }i, k\in\{1,\ldots,m\}\}
    \nonumber\\
    &\geq (1-n^\beta e^{-\Delta(\delta_0)n^{1-\beta}})
    (1-m^2 e^{-(1-e^s+s(1+\delta_0))l/m})
    \nonumber\\
    &=(1-n^\beta e^{-\Delta(\delta_0)n^{1-\beta}})
    (1-e^{2\beta \log n -n^{\gamma-\beta}(1-e^s+s(1+\delta_0))}),
    \nonumber
\end{align}
which converges to one as $n$ goes to infinity, by choosing
$s=\log(1+\delta_0)$.

3) Regimes B-1, B-2, and B-3 $(\beta+2\gamma> 1 \textrm{~and~}
\frac{1}{3}\leq\beta<1)$: Since $\beta+2\gamma>1$, the number of
simultaneously transmitting nodes in each cell is
$(n/m)^{1/2-\epsilon}$. From Lemma \ref{Lem:NumPairs}, we set
$a=\log n$ because $\frac{1}{3}\leq\beta<1$. Then, we have
\ifx \doubleColumn \undefined 
\begin{align}\label{Eq:RegimeG-ProbB} 
    \Pr\left\{\sum_{j=1}^{(n/m)^{1/2-\epsilon}}B_j\geq\log
    n\right\}
    &=\Pr\left\{e^{s\sum_{j=1}^{(n/m)^{1/2-\epsilon}}B_j}\geq
    e^{s\log n}\right\}
    \leq (\mathbb{E}[e^{sB_1}])^{(n/m)^{1/2-\epsilon}}e^{-s\log n}
    \nonumber\\
    &=\left(e^s\frac{1}{m}+\left(1-\frac{1}{m}\right)\right)^{(n/m)^{1/2-\epsilon}}e^{-s\log n}
    \nonumber\\
    &\leq e^{-(1/m)(1-e^s)(n/m)^{1/2-\epsilon}}e^{-s\log n}
    =e^{-(n/m^3)^{1/2-\epsilon'}(1-e^s)-s\log n},
\end{align}
\else
\begin{align}\label{Eq:RegimeG-ProbB} 
    &\Pr\left\{\sum_{j=1}^{(n/m)^{1/2-\epsilon}}B_j\geq\log
    n\right\} \nonumber\\
&    =\Pr\left\{e^{s\sum_{j=1}^{(n/m)^{1/2-\epsilon}}B_j}\geq
e^{s\log n}\right\}
    \nonumber\\
    &\leq (\mathbb{E}[e^{sB_1}])^{(n/m)^{1/2-\epsilon}}e^{-s\log n}
    \nonumber\\
    &=\left(e^s\frac{1}{m}+\left(1-\frac{1}{m}\right)\right)^{(n/m)^{1/2-\epsilon}}e^{-s\log n}
    \nonumber\\
    &\leq e^{-(1/m)(1-e^s)(n/m)^{1/2-\epsilon}}e^{-s\log n}
    \nonumber\\
    &
    =e^{-(n/m^3)^{1/2-\epsilon'}(1-e^s)-s\log n},
\end{align}
\fi where $\epsilon'>0$ is an arbitrarily small constant and $s$
is a positive real value. Using (\ref{Eq:RegimeG-ProbB}), the
probability in (\ref{Eq:ProbXki}) is given by
\begin{align}
    &\Pr\{X_{ki}<\log n\textrm{ for all }i, k\in\{1,\ldots,m\}\}
    \nonumber\\
    &\geq (1-n^\beta e^{-\Delta(\delta_0)n^{1-\beta}})
    (1-m^2 e^{-(n/m^3)^{1/2-\epsilon'}(1-e^s)-s\log n})
    \nonumber\\
    &=(1-n^\beta e^{-\Delta(\delta_0)n^{1-\beta}})
    (1-e^{2\beta \log n -n^{(1-3\beta)/2-\epsilon''}(1-e^s)-s\log
    n}), \nonumber
\end{align}
which converges to one as $n$ goes to infinity, by choosing
$s=(2+\delta_0)\beta$.

4) Regime B-4 $(\beta+2\gamma> 1 \textrm{~and~}
0\leq\beta<\frac{1}{3})$: Since $\beta+2\gamma>1$, the number of
simultaneously transmitting nodes in each cell is
$(n/m)^{1/2-\epsilon}$. From Lemma \ref{Lem:NumPairs}, we set
$a=(1+\delta_0)^2(n/m^3)^{1/2-\epsilon'}$ because
$0\leq\beta<\frac{1}{3}$. Then, we have
\ifx \doubleColumn \undefined 
\begin{align} \label{Eq:RegimeH-ProbB} 
    &\Pr\left\{\sum_{j=1}^{(n/m)^{1/2-\epsilon}}B_j\geq (1+\delta_0)^2\left(\frac{n}{m^3}\right)^{1/2-\epsilon'}\right\}
    \nonumber\\
    &=\Pr\left\{e^{s\sum_{j=1}^{(n/m)^{1/2-\epsilon}}B_j}\geq
    e^{s(1+\delta_0)^2(n/m^3)^{1/2-\epsilon'}}\right\}
    \leq (\mathbb{E}[e^{sB_1}])^{(n/m)^{1/2-\epsilon}}e^{-s(1+\delta_0)^2(n/m^3)^{1/2-\epsilon'}}
    \nonumber\\
    &=\left(e^s\frac{1}{m}+\left(1-\frac{1}{m}\right)\right)^{(n/m)^{1/2-\epsilon}}e^{-s(1+\delta_0)^2(n/m^3)^{1/2-\epsilon'}}    \nonumber\\
    &\leq e^{-(1/m)(1-e^s)(n/m)^{1/2-\epsilon}}e^{-s(1+\delta_0)^2(n/m^3)^{1/2-\epsilon'}}
    =e^{-(n/m^3)^{1/2-\epsilon''}(s(1+\delta_0)^2+1-e^s)},
\end{align}
\else
\begin{align} \label{Eq:RegimeH-ProbB} 
    &\Pr\left\{\sum_{j=1}^{(n/m)^{1/2-\epsilon}}B_j\geq (1+\delta_0)^2\left(\frac{n}{m^3}\right)^{1/2-\epsilon'}\right\}
    \nonumber\\
    &=\Pr\left\{e^{s\sum_{j=1}^{(n/m)^{1/2-\epsilon}}B_j}\geq
    e^{s(1+\delta_0)^2(n/m^3)^{1/2-\epsilon'}}\right\} \nonumber\\
    &\leq (\mathbb{E}[e^{sB_1}])^{(n/m)^{1/2-\epsilon}}e^{-s(1+\delta_0)^2(n/m^3)^{1/2-\epsilon'}}
    \nonumber\\
    &=\left(e^s\frac{1}{m}+\left(1-\frac{1}{m}\right)\right)^{(n/m)^{1/2-\epsilon}}e^{-s(1+\delta_0)^2(n/m^3)^{1/2-\epsilon'}}    \nonumber\\
    &\leq e^{-(1/m)(1-e^s)(n/m)^{1/2-\epsilon}}e^{-s(1+\delta_0)^2(n/m^3)^{1/2-\epsilon'}}
    \nonumber\\
    &=e^{-(n/m^3)^{1/2-\epsilon''}(s(1+\delta_0)^2+1-e^s)},
\end{align}
\fi where $\epsilon''>0$ is an arbitrarily small constant and $s$
is a positive real value. Using (\ref{Eq:RegimeH-ProbB}), the
probability in (\ref{Eq:ProbXki}) is given by
\begin{align}
    &\Pr\{X_{ki}<(1+\delta_0)^2(n/m^3)^{1/2-\epsilon'}\textrm{ for all }i, k\in\{1,\ldots,m\}\}
    \nonumber\\
    &\geq (1-n^\beta e^{-\Delta(\delta_0)n^{1-\beta}})
    (1-m^2 e^{-(n/m^3)^{1/2-\epsilon'}(s(1+\delta_0)^2+1-e^s)})
    \nonumber\\
    &=(1-n^\beta e^{-\Delta(\delta_0)n^{1-\beta}})
    (1-e^{2\beta \log n
    -n^{(1-3\beta)/2-\epsilon''}(s(1+\delta_0)^2+1-e^s)}), \nonumber
\end{align}
which converges to one as $n$ goes to infinity, by choosing
$s=\log(1+\delta_0)$.

From the analysis above, we obtain the number of S--D pairs
between two BSs in (\ref{Eq:BoundPacketsIMH}). Since the
transmission rate of each source is $T_{n,\textrm{IMH}}/n$ and
$\min\{l,(n/m)^{1/2-\epsilon}\}$ nodes among $n/m$ source nodes in
each cell simultaneously transmit their packets, the rate of each
backhaul link between two BSs is given by
$\frac{T_{n,\textrm{IMH}}}{n}\frac{n}{m}\frac{1}{\min\{l,(n/m)^{1/2-\epsilon}\}}$.
Therefore, the required rate is given by
\ifx \doubleColumn \undefined 
\begin{align}\label{Eq:CBS-IMH-Rate-Tn} 
    C_{\textrm{BS,IMH}}
    = \left\{ \begin{array}{ll}
    \Omega\left(\frac{T_{n,\textrm{IMH}} }{ml}\log n\right)
    =\Omega(\log n)
    &\textrm{for Regime A-1}\\
    \Omega\left(\frac{T_{n,\textrm{IMH}} }{ml}\frac{l}{m}\right)
    =\Omega(\frac{l}{m})
    &\textrm{for Regime A-2}\\
    \Omega\left(\frac{T_{n,\textrm{IMH}}}{m(n/m)^{1/2-\epsilon}}\log n\right)
    =\Omega\left(\log n\right)
    &\textrm{for Regimes B-1, B-2, and B-3}\\
    \Omega\left(\frac{T_{n,\textrm{IMH}}}{m(n/m)^{1/2-\epsilon}}n^{-\epsilon}\sqrt{\frac{n}{m^3}}\right)
    =\Omega\left(n^{-\epsilon}\sqrt{\frac{n}{m^3}}\right)
    &\textrm{for Regime B-4},
    \end{array} \right.
\end{align}
\else
\begin{align}\label{Eq:CBS-IMH-Rate-Tn} 
    &C_{\textrm{BS,IMH}}
    \nonumber\\
    &= \left\{ \begin{array}{ll}
    \Omega\left(\frac{T_{n,\textrm{IMH}} }{ml}\log n\right)
    =\Omega(\log n)
    \\\textrm{for Regime A-1}\\
    \Omega\left(\frac{T_{n,\textrm{IMH}} }{ml}\frac{l}{m}\right)
    =\Omega(\frac{l}{m})
    \\\textrm{for Regime A-2}\\
    \Omega\left(\frac{T_{n,\textrm{IMH}}}{m(n/m)^{1/2-\epsilon}}\log n\right)
    =\Omega\left(\log n\right)
    \\\textrm{for Regimes B-1, B-2, and B-3}\\
    \Omega\left(\frac{T_{n,\textrm{IMH}}}{m(n/m)^{1/2-\epsilon}}n^{-\epsilon}\sqrt{\frac{n}{m^3}}\right)
    =\Omega\left(n^{-\epsilon}\sqrt{\frac{n}{m^3}}\right)
    \\\textrm{for Regime B-4},
    \end{array} \right.
\end{align}
\fi which completes the proof of Lemma~\ref{Lem:CBS-IMH}.


\end{document}